%% file: main.tex
\documentclass[aps,physrev,twocolumn, longbibliography, nofootinbib, superscriptaddress, nobibnotes, floats]{revtex4-2}
\usepackage{amssymb, amsmath, graphicx, dcolumn, color,units, xspace, mathtools, physics, tensor, bm, lipsum, revsymb, xcolor}
\usepackage{enumitem}
\usepackage[normalem]{ulem}
\usepackage{caption}
\usepackage{hyperref}
\usepackage{booktabs,dcolumn, subcaption}
\usepackage{ragged2e, tikz}
\usetikzlibrary{arrows.meta,positioning,fit,backgrounds,calc}
\tikzset{
  font=\footnotesize,
  proc/.style ={draw,rounded corners=3pt,fill=gray!8,align=center,inner sep=5pt,minimum height=9mm},
  dgdata/.style={draw,fill=blue!8,align=center,inner sep=4pt,minimum height=8mm},
  store/.style ={draw,fill=cyan!12,align=center,inner sep=4pt},
  dgout/.style ={draw,very thick,rounded corners=2pt,fill=yellow!20,align=center,inner sep=6pt},
  ar/.style    ={-{Stealth[length=2.4mm]},semithick},
}

\newcommand{\emptyplaceholder}[1]{#1}
\newcommand{\paperchanges}[1]{#1}

\begin{document}

\title{\texttt{BilbyFlow}: user-friendly neural posterior estimation for gravitational-wave astronomy}

\author{Liam Pinchbeck}
\affiliation{School of Physics and Astronomy, Monash University, Clayton, Victoria 3800, Australia}

\author{Eric Thrane}
\affiliation{School of Physics and Astronomy, Monash University, Clayton, Victoria 3800, Australia}
\affiliation{OzGrav: The ARC Centre of Excellence for Gravitational-Wave Discovery, Clayton, Victoria 3800, Australia}

\author{Csaba Balazs}
\affiliation{School of Physics and Astronomy, Monash University, Clayton, Victoria 3800, Australia}

\author{Paul D. Lasky}
\affiliation{School of Physics and Astronomy, Monash University, Clayton, Victoria 3800, Australia}
\affiliation{OzGrav: The ARC Centre of Excellence for Gravitational-Wave Discovery, Clayton, Victoria 3800, Australia}

\begin{abstract}
    Bayesian inference plays a central role in the new field of gravitational-wave astronomy. 
    However, traditional Bayesian inference with stochastic samplers is computationally expensive, taking hours to days per event. 
    Transformative changes are therefore required to enable the science of next-generation observatories whose event rates and signal-to-noise ratios will increase significantly over the current generation. 
    Recent work has shown that neural posterior estimation (NPE) is a promising path forward.
    A neural net is trained to approximate the posterior distribution of gravitational-wave parameters, allowing generation of posterior samples in a fraction of the time required by stochastic samplers. 
    In this work, we introduce \texttt{BilbyFlow}, which harnesses the power of NPE in the popular \texttt{Bilby} code suite. 
    We use \texttt{BilbyFlow} to analyze a subset of 38 high-mass events from the third LIGO-Virgo-KAGRA Gravitational-Wave Transient Catalog (GWTC-3).
    For 29 events (76\%), we obtained an importance-sampling efficiency $>$1\%, allowing us to produce reliable posterior distributions within \emptyplaceholder{\unit[3]{min}-\unit[1.5]{hours}}.
    For the other events, with importance-sampling efficiency $\ll$1\%, the run time can be as long as \emptyplaceholder{\unit[35]{hours}}. 
    We achieve a median importance-sampling efficiency of \emptyplaceholder{7}\%, which is roughly comparable to the \texttt{DINGO} package.
    We aim to significantly improve this efficiency with further development to make the runtime more reliably ${\cal O}(\text{min})$. 
    \texttt{BilbyFlow} is open source and \texttt{pip}-installable.
\end{abstract}

\maketitle

\section{Introduction}
Parameter estimation is one of the central tasks of gravitational-wave astronomy.  
Posterior sample estimates of binary parameters underpin a wealth of downstream analyses including population studies which probe the astrophysics of binaries and massive stars \cite[e.g.,][]{gwtc-5_pop}, cosmological measurements of the expanding Universe \cite[e.g.,][]{gwtc-5_cosmo}, and tests of general relativity \cite[e.g.,][]{gwtc-4_tgr}. 
In practice, parameter estimation is typically performed using stochastic sampling methods such as nested sampling \cite{Skilling} and Markov Chain Monte Carlo \cite{Metropolis,Hastings}.  These methods are robust and accurate, and provide the gold-standard results for gravitational-wave inference.

However, the computational cost of stochastic sampling is substantial.  A typical compact-binary analysis requires hours to days of computation depending on the waveform model, dimensionality of the parameter space, and structure of the posterior distribution.  More challenging events---e.g., with long durations \cite[e.g.,][]{GW170817} or signs of eccentricity \cite[e.g.,][]{Morras}---can require significantly longer runtimes.  The cost arises because posterior evaluation requires repeated likelihood calculations, each involving waveform generation and comparison with detector data.  Consequently, obtaining a sufficiently large set of posterior samples can require millions of likelihood evaluations.

The problem will become even worse with the arrival of next-generation observatories like Cosmic Explorer~\citep{CE,Reitze} and the Einstein Telescope~\citep{ET}.
With improved sensitivity and wider observing band, future observatories will measure binaries for longer durations (up to 90~minutes) with far greater signal-to-noise ratios, which both increase the computational costs of stochastic samplers.
Various tricks are employed to speed up the likelihood calculation, including reduced-order methods \cite{Smith,Canizares,Field,Baker}, heterodyning / relative binning \cite{Cornish,Zackay}.
However, even using these tricks, it currently takes 128~CPU cores approximately 48~hours to analyze one loud binary neutron-star event \citep{nir_roq}.

The problem is exacerbated by the increasing detection rate, which is projected to reach ${\cal O}(\unit[1]{min^{-1}})$ with next-generation observatories \cite{GW170817_stoch}.
At the same time, waveform models are becoming more sophisticated, and scientific analyses are becoming more ambitious.  Together, these developments place increasing pressure on traditional inference pipelines. New approaches are therefore required that retain the accuracy of Bayesian inference while reducing its computational cost.

Recent advances in machine learning have provided a promising alternative to traditional stochastic inference.  Several groups have demonstrated that amortized\footnote{The word ``amortized'' comes from the world of finance. 
It means ``spread out over many uses.'' 
In this context, it means that a neural net that is trained once can be applied to a variety of situations.} neural density estimation can accurately approximate gravitational-wave posteriors while reducing the computational cost of inference by orders of magnitude \cite{Green_2020,Dax_2021,Dax_2023,Wildberger_2023, LABRADOR_ref, Beta_Flows_Ref}.  
These simulation-based inference (SBI) methods train neural networks on large catalogs of simulated gravitational-wave signals before they are applied to observational data.
The key feature of this approach is that the expensive optimization is performed during training rather than inference.  Once trained, the network directly predicts an approximation to the posterior distribution conditioned on the observed data. 
Posterior samples can then be generated in less than a second using a simple forward pass through the network.

Until now, however, neural density estimation has not been available as part of the \texttt{Bilby} code suite \citep{Bilby,BilbyValidation}, which is widely used for gravitational-wave inference.
We introduce \texttt{BilbyFlow}, which addresses this need by bringing neural posterior estimation to the \texttt{Bilby} software framework.  \texttt{Bilby} is one of the most widely used inference packages in gravitational-wave astronomy and forms the basis of many analyses performed within the LIGO-Virgo-KAGRA (LVK) Collaboration \cite{LIGOScientific, VIRGO2014yos, KAGRA2020} and the wider community.  By building directly on \texttt{Bilby}, \texttt{BilbyFlow} is immediately compatible with its existing infrastructure for waveform generation, detector modelling, priors, likelihoods, and data handling.

\texttt{BilbyFlow} inherits the software design principles that have made \texttt{Bilby} popular.  The framework is modular, extensible, and straightforward to use.  It benefits from the continued development and maintenance of the broader \texttt{Bilby} community, ensuring that new waveform models, likelihoods, and infrastructure improvements become available to \texttt{BilbyFlow} users.  In this way, \texttt{BilbyFlow} combines the speed of neural posterior estimation with the flexibility and maturity of the \texttt{Bilby} ecosystem.

The remainder of this manuscript is organized as follows.
In Section~\ref{sec:method}, we describe our methodology.
In Section~\ref{sec:Results}, we demonstrate \texttt{BilbyFlow} on synthetic, simulated data and events from LIGO--Virgo-KAGRA's third Gravitational-Wave Transient Catalog~\cite[GWTC-3;][]{GWTC3_paper}.
We show that \texttt{BilbyFlow} can produce results consistent with \texttt{Bilby} and assess the efficiency with which it produces true posterior samples.
In Section~\ref{sec:discussion} we provide concluding remarks and discuss the future developments planned for \texttt{BilbyFlow}.

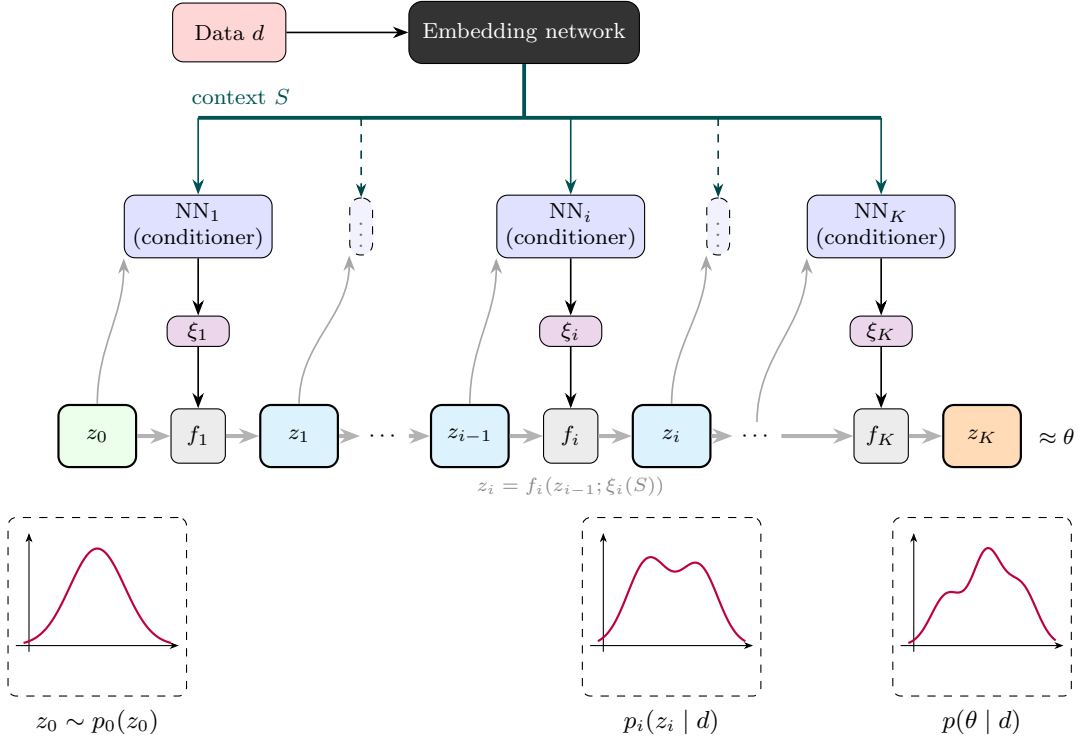
\begin{figure*} 
    \centering
    \resizebox{0.8\linewidth}{!}{\input{Tables_and_Diagrams/general_conditional_flow_architecture}}
    \caption{A conditional normalizing flow diagram showing the structure of a general neural posterior estimation workflow. The observed data $d$ (in our case, the gravitational-wave strain information) is compressed by an embedding network into a context vector $S$, which is given to every transform conditioner network in the flow. At each normalizing flow transformation layer $i$, a dedicated conditioner network $\mathrm{NN}_i$ receives $S$ together with the current latent $\vec{z}_{i-1}$ and outputs the transform parameters $\xi_i$ (which can be thought of as a function of the embedding context $S$).
    The bijection $f_i(\vec{z}_{i-1} \vert \xi_i$) then maps $\vec{z}_{i-1}$ to $\vec{z}_i$. Composing $K$ of these transforms maps the base distribution $p_0(\vec{z}_0)$ (typically a standard normal) to the posterior $p(\theta\vert d)$. The embedding network and each conditioner have independent learnable parameters; the transforms $f_i$ are static/parameter-free. During training, the transformation sequence is evaluated in the inverse or \emph{normalizing} direction $\theta\rightarrow\vec{z}_0$ to compute the flow density.
    During inference the samples are generated in the opposite direction $\vec{z}_0\rightarrow \theta$ \emph{forward} through the transformation sequence.}
    \label{fig:SBIFlowDiagram}
\end{figure*}

\section{Methodology}\label{sec:method}
\subsection{Goals}
The posterior distribution for gravitational-wave astronomy is given by
\begin{align}
    p(\theta | d) = \frac{1}{{\cal Z}(d)} \,
    {\cal L}(d | \theta) \, 
    \pi(\theta) .
\end{align}
Here, $\theta$ are typically the fifteen parameters describing a binary black hole system: seven \textit{extrinsic} parameters describing how the binary is situated in spacetime relative to Earth and eight \textit{intrinsic} parameters involving the mass and spins of the two compact objects.\footnote{In this work, we focus on quasi-circular inspiral of black hole binaries, which contain the fifteen parameters mentioned. The framework we develop is simply extendable to include, e.g., eccentricity parameters and/or tides in the case of neutron star binaries.}
Meanwhile, $d$ is the data, which consists of a strain time series or, equivalently, a frequency series.
The quantity ${\cal Z}(d)$ is the Bayesian evidence, which serves here as a normalization constant.
The likelihood is denoted ${\cal L}(d|\theta)$ and the prior is given by $\pi(\theta)$.
For a review of Bayesian inference in gravitational-wave astronomy, see Ref.~\cite{intro}.

Our primary goal in this paper is to efficiently draw samples from the posterior distribution $p(\theta | d)$, which characterize the binary parameters.
These posterior distributions sometimes yield important discoveries, as was the case for GW231123---an unusually massive binary (total mass $\gtrsim 240 M_\odot$) with at least one rapidly spinning black hole~\cite{GW231123}.
The samples are also an essential ingredient in population studies for analyzing the distributions of binary parameters; \cite[see, e.g.,][]{gwtc-5_pop}.

A secondary goal is to efficiently estimate the Bayesian evidence, which is used for various model-selection studies:
\begin{align}\label{eq:evidence}
    {\cal Z}(d) \approx \frac{1}{n}\sum_{k=1}^n 
    \frac{
    {\cal L}(d|\theta_k)\,
    \pi(\theta_k)
    }{q_{\varphi}(\theta_k|d)} .
\end{align}
Here, the sum is over draws from $q_{\varphi}(\theta_k|d)$---the approximate posterior obtained with neural posterior estimation (NPE).\footnote{Equation \ref{eq:evidence} is an example of \textit{importance sampling}.}
Rapid estimation of the Bayesian evidence could be a transformative tool, facilitating detection of the stochastic background \cite{TBS,TBS-PE,tbs_mdc,bers}, deep searches for sub-threshold signals \cite{Veitch2010,BCR,Pratten2021}, and ``model criticism'' studies that are not currently practical due to computational constraints \cite{Lstroke, polanska2024acceleratedbayesparamest, Srinivasan_2024}.
The remainder of this Section is about how \texttt{BilbyFlow} obtains posterior samples with NPE.

\subsection{Flow basics}
The crucial component of our SBI framework is the normalizing flow. 
Normalizing flows are composed of a set of typically simple transformations\footnote{Although more expressive transformations exist, e.g. \cite{decao2019blockneuralautoregressiveflow, wehenkel2021UNAF}.} $f_i$ with parameters $\xi_i$ dictated by dedicated neural networks for each transform ($\textrm{NN}_i$) \cite{rezende_VI_with_NF, papamakarios_NF_for_prob_modeling}; see Fig.~\ref{fig:SBIFlowDiagram}.\footnote{In this context, a ``simple transformation'' is an invertible mapping that is easy to evaluate with a computationally cheap Jacobian.} 
We denote the combined set of weights for the neural networks with $\varphi$. 
These transformations map a base distribution (of the same dimension as the parameters of interest $\theta$) to a target distribution, in our case, the posterior. 

The loss function for neural density estimators such as normalizing flows are typically formulated as the KL-divergence between the \textit{proposal distribution} described by the flow $q_\varphi(\theta)$, and the \textit{target distribution} $p(\theta | d)$. 
In non-data-amortized settings---that is to say, using a model trained for a specific dataset---we use the reverse KL-divergence estimated via Monte-Carlo samples from $q_\varphi(\theta)$, $\text{KL}(q_\varphi(\theta) \parallel \pi(\theta\vert d))$ (see e.g. \cite{Mould_2025, wolfe2026neuralbayesianupdatespopulations}). 
The approximate distribution does not depend explicitly on the data $d$, as it is not amortized with respect to the data. 
One cannot give it a new set of data and get a posterior density.

In SBI we instead use the expected \emph{forward} KL-divergence\footnote{We are careful here to distinguish between the forward and the reverse KL-divergence. The two are not the same because KL-divergence is not a \emph{distance}. All distances are divergences, but only symmetric divergences can be distances and the KL-divergence is not symmetric.} which is calculated with respect to the joint distribution $\pi(\theta, d)=\mathcal{L}(d|\theta)\pi(\theta)$:
\begin{align}&\mathbb{E}_{\pi(d)}\left[\text{KL}(\pi(\theta\vert d) \parallel q_\varphi(\theta|d) ) \right] \nonumber
    \\&= \int_{d} \int_{\theta} \pi(\theta, d) \log\frac{\pi(\theta\vert d) }{q_\varphi(\theta|d) }  d\theta d(d) \nonumber \\
    &\approx \frac{1}{N_S} \sum_{(\theta^{(i)}, d^{(i)}) \sim \pi(\theta, d) }^{N_S} \log \pi(\theta^{(i)}\vert d^{(i)}) - \log q_\varphi(\theta^{(i)}|d^{(i)}).
\end{align}
Here, $d(d)$ refers to an infinitesimal change in the data and $\mathbb{E}_{\pi(d)}$ refers to the expectation value averaged over draws from $\pi(d)$.

This alleviates the need for explicit target densities, only requiring paired samples of the data  and the parameters $(d^{(i)}, \theta^{(i)})$ for training. Then, because the sum can be split, and neither the samples nor the density in the first term are conditional on $\varphi$, the first term is constant under optimization. Hence, the objective for fitting the flow is typically given as
\begin{align}
    L(\varphi) = -\sum_{(\theta^{(i)}, d^{(i)}) \sim \pi(\theta, d) }^{N_S} \log q_\varphi(\theta^{(i)}|d^{(i)}).
\end{align}

The flow has an explicit dependence on $d$ because it is amortized with respect to the data; the embedding and conditioner networks learn to map information in the realization of $d$ to the corresponding posterior over $\theta$. 
In practice, this means that SBI has a large upfront cost (e.g., training the normalizing flow and embedding network) but, having paid for the training, the cost to apply to data is cheap.
The performance of SBI can be competitive or even superior when compared to traditional methods \cite{frazier2024statisticalaccuracyneuralposterior}. 

During inference, the time it takes to generate samples from the approximate posterior $q_\varphi(\theta|d)$ is just the time it takes for the forward pass through the embedding network and neural networks dictating the parameterization of the transforms, and then for the samples from the noise distribution $\vec{z}_0$ to be fed through these transforms.
For the specific case of gravitational-wave inference, it is possible to generate thousands of samples from $q_\varphi(\theta|d)$ in $\lesssim\unit[1]{s}$.
The workflow is shown in Fig.~\ref{fig:SBIFlowDiagram}.

\subsection{Importance sampling}
If the approximate posterior $q_\varphi(\theta |d)$ provided a nearly perfect match for the actual posterior $p(\theta |d)$, we could treat the draws from $q_\varphi$ as bona fide posterior samples.
Unfortunately, subtle differences between $q_\varphi(\theta|d)$ and $p(\theta|d)$ mean that these two distributions are measurably different---even if they have qualitatively similar corner plots.
We therefore distinguish between \textit{proposal samples} drawn from $q_\varphi(\theta|d)$ and \textit{posterior samples} drawn from $p(\theta|d)$.
See, e.g., Fig.~\ref{fig:BilbyFlow_prior_vs_BilbyFlow_full} in the Appendix, which shows the credible intervals for the proposal samples in orange and the credible intervals obtained with posterior samples in blue.

The difference between proposal samples and posterior samples can be measured using the importance sampling efficiency \cite{Kish_1965}, which is defined as \cite[see, e.g.,][]{Payne2019}:
\begin{align}
    \epsilon \equiv 
    \frac{1}{n}
    \frac{\left(\sum_k^n w_k\right)^2}{\sum_k^n w_k^2} ,
\end{align}
where $n$ is the number of proposal samples and
\begin{align}
    w_k \propto \frac{p(\theta_k | d)}{q_{\varphi}(\theta_k | d)} ,
\end{align}
are weights comparing the target distribution $p(\theta|d)$ with the proposal distribution $q_\varphi(\theta|d)$.
In practice, we do not have an expression for $p(\theta|d)$ because we do not know the Bayesian evidence ${\cal Z}$.
But, since we only care about the relative value of weights, we can just leave off the evidence so that
\begin{align}\label{eq:weight}
    w_k = \frac{{\cal L}(d|\theta_k)\,\pi(\theta_k)}{q_{\varphi}(\theta_k | d)} .
\end{align}

The efficiency tells us how close the proposal distribution $q_\varphi(\theta|d)$ has gotten to the target distribution $p(\theta|d)$.
If the efficiency is reasonably high $\gtrsim 1\%$, we can use importance sampling to generate posterior samples from the proposal samples.
The simplest way to do this is to simply assign the weight from Eq.~\ref{eq:weight} to each proposal sample.
The weighted proposal samples are posterior samples.
However, by importance sampling, we get fewer effective posterior samples $n_\text{eff}$ than the number of proposal samples that we started with $n$:
\begin{align}
    n_\text{eff} = \epsilon \,
    n
\end{align}
For example, if we want $10^4$ posterior samples, we need approximately $2\times10^4$ proposal samples from a flow with $\epsilon = 50\%$ in order to get the right number of actual posterior samples after importance sampling.

In practice, we consider a sampling efficiency $\gtrsim 1\%$ as ``good.''
Even if one needs to generate 100 flow samples for every one posterior sample, the flow still provides a practical method for rapidly generating posterior samples.
If the efficiency is $\ll 1\%$, one can still in principle use the flow to obtain posterior samples, but the calculation may start to become as time-consuming as traditional inference methods.
We therefore set a goal $\epsilon \geq 1\%$ for \texttt{BilbyFlow}.

\subsection{Rejection sampling}
Historically, some users have expressed a preference for posterior samples with equal weights.
Indeed, most downstream analyses that rely on posterior samples (e.g., \texttt{GWPopulation} \cite{gwpop,gwtc-5_pop}) assume equal weights.
The weighted samples created by \texttt{BilbyFlow} can be converted into equal-weight samples by rejection sampling.
Rejection sampling, if it is required, must be carried out as a post-processing step after the generation of weighted samples.
However, rejection-sampling efficiency is roughly ten times less than importance-sampling efficiency.
Thus, equal-weight posterior samples obtained by rejection sampling are computationally far more expensive to produce than weighted samples.

It is already challenging to achieve reliably high importance-sampling efficiency for the vast majority of gravitational-wave events.
We therefore suggest that the community would be well served by adapting existing pipelines to ingest importance-sample weights.
This will make it easier to achieve rapid inference for a large percentage of events.

To that end, when reporting values, we refer to the weighted sample results.
Even for weighted samples, the reweighting post-processing step is far more computationally expensive than the generation of the proposal samples.
Thus, the computation time for NPE is limited by the time it takes to ``fix'' the imperfect modelling of the posterior distribution with a normalizing flow.

\subsection{Subtleties}\label{sec:subtleties}
Since the importance sampling is the main computational bottleneck in NPE, we employ tricks to make this step as fast as possible.
During training, we treat the phase of coalescence $\phi_c$, polarization angle $\psi$ and the time of coalescence $t_c$ as nuisance parameters.
That is, we randomize their values in our training set, but we do not teach the flow to learn the posterior distribution for these three parameters.
This makes the training easier because we do not have to model the complicated correlations between $\phi_c,\psi, t_c$ and the other parameters.\footnote{The DINGO pipeline handles this by employing group equivariant posterior estimation (GNPE), but we choose instead to implicitly marginalize.}
However, by making the training easier, we make the post-processing harder.
The numerator of our weights must include marginalization over $\phi_c, \psi, t_c$:
\begin{align}
    w_k = \frac{\int d\phi_c \int d\psi \int dt_c\, {\cal L}(d|\theta_k, t_c, \phi_c) \, \pi(\theta) \, \pi(\phi_c) \, \pi(\psi)\, \pi(t_c)}{q_\varphi(\theta_k|d)}.
\end{align}
(In this subsection, $\theta$ represents all the binary parameters except $\phi_c, \psi, t_c$.)
This marginalization \paperchanges{(performed numerically, see Appendix~\ref{sec:prior_swap} for more details)} ensures that the numerator and denominator have the same number of parameters.

We do the integral over $t_c$ with a fast Fourier transform using the method described in Ref.~\cite{intro}; see their Appendix~C1.
We include the \texttt{Bilby} ``jitter'' term to probe $t_c$ values in between the grid defined by the fast Fourier transform.

\subsection{Flow details}
To make the conditional flow as widely applicable as possible, we define our gravitational-wave priors to include as many events as possible without making the training step too difficult. 
As a first step, we focus on signals that fit within \unit[4]{s} segments, which limits the domain of utility to events with chirp mass $\gtrsim 10 M_\odot$.
By focusing on these shorter signals (which make up the majority of LVK detections) we are able to manage GPU memory usage and computation time. 
The framework is additionally restricted to two-detector signals from the LIGO Hanford and Livingston sites as the neural network currently requires a fixed input. 
If an extra detector is available, its information is not used.
(Of course, this will be addressed in a future update.)
The priors used for training are shown in Table~\ref{tab:FlowTrainingPriors} of Appendix~\ref{sec:TrainingAppendix}.

It is also helpful to restrict the luminosity distance prior to make training easier; see~\cite{Dax_2021}. 
The high signal-to-noise ratio (SNR) regime---corresponding to low luminosity distances---is where the posterior is most narrow and diverges most from the prior.
However, that is also the region of parameter space that receives the least samples according to the physically motivated prior: $\pi(d_L)\propto d_L^2$. 
Hence, a log-uniform prior is used on top of a training curriculum \cite{Bengio_Curriculum_Learning} in order to slowly introduce events with larger $d_L$ over the course of training.\footnote{See Ref.~\cite{wang2021surveycurriculumlearning} for some examples of curriculum learning.}
This ensures that more samples populate the high-SNR region of parameter space and salient features of the waveform morphology are learned before having to cope with the varying noise floor. 
We reweight the density and samples after the training run to recover the physical prior.

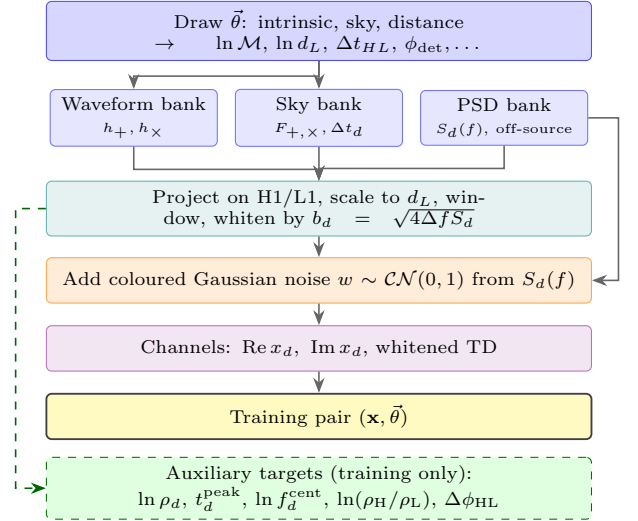
\begin{figure}
  \centering
  \makebox[\linewidth][c]{\input{Tables_and_Diagrams/datagen_diagram_plain_english}}
  \caption{Simulation pipeline for generating training data for the conditional flow framework.
  For each sampled parameter vector $\theta$, an \texttt{IMRPhenomXPHM} waveform is generated at a reference luminosity distance ($d_\mathrm{ref}=\unit[1]{Mpc}$), projected onto the H1--L1 network with sky-sampled antenna responses $(F_+, F_\times)$ and time delays (the detector-frame coordinates $\Delta t_{HL}, \varphi_{\mathrm{det}}$ are stored as inference targets), rescaled to the sampled distance ($d_\mathrm{ref}/d_L$), and Tukey-windowed.
  Power spectral densities $S_d(f)$ are estimated from off-source detector segments and define the whitening factor $b_d$.
  During training only, we use physically interpretable summaries of the noiseless whitened signal ($\rho_d, \phi_d, t^{\mathrm{pk}}_d, f_{\mathrm{cent}}, \Delta t_{HL}, \dots$) as auxiliary regression targets to shape the embedding.}
  \label{fig:SimulationPipeline}
\end{figure}

We feed prior samples into \texttt{Bilby}'s standard simulator workflow as detailed in Fig.~\ref{fig:SimulationPipeline}. 
The key consideration is that the noise, extrinsic sky parameters, and intrinsic waveform parameters can be simulated independently. 
This means that a large number of each can be simulated, and then combined randomly, increasing the effective number of samples combinatorially for training compared to generating all components jointly.

With the simulated training pairs, we construct a conditional normalizing flow to model the conditional density $p(\theta\vert d)$. 
The strain data is then compressed into a 512-dimensional embedding via four convolution layers followed by ResNet-18 encoders \cite{He2016}.
We additionally condition on estimated noise power spectral densities (PSDs) derived from off-segment data to make the setup truly amortized to different observation time periods.\footnote{If we were to train on a single PSD realization, then the conditioning would not be necessary, but would also mean the setup could only be applied to data segments with the same noise distribution.} 
For real data, the noise PSD is estimated from surrounding segments.
For simulated data, they are the PSDs drawn from the noise bank used to generate each training sample.
The estimated PSD is standardized and then fed into a multi-layer perceptron (MLP) to create a low-dimensional representation.
These two representations are then fed into the neural networks parameterizing the flow transform parameters.
By training on the \textit{estimated} PSD, we additionally implicitly condition on the PSD uncertainty model {\`a} la Refs.~\cite{Talbot2020,Biscoveanu2020}.
We expect that the posterior approximations should be slightly broader than those obtained with a standard \texttt{Bilby} run, decreasing the importance-sampling efficiency but ensuring that the reweighted samples are unbiased. 

\begin{figure*}[t]
  \centering
  \makebox[0.8\textwidth][c]{\input{Tables_and_Diagrams/flow_architecture}}
  \caption{Architecture of the neural posterior estimator in this work. 
The network input $d$ comprises the whitened H1--L1 strain---represented in the frequency and time domains---together with a per-detector log-PSD context block. 
The strain is encoded by two parallel branches, one per representation, each a 1-D convolutional stem followed by a ResNet-18; their outputs are concatenated and passed through an MLP head to a $512$-dimensional strain embedding. 
The PSD context is encoded by a separate MLP to a $64$-dimensional vector, and the two are concatenated into the $576$-dimensional conditioning context $\mathbf{h}$.
A conditional neural spline flow maps the twelve inferred parameters $\vec\theta$ to a standard normal base density with the concatenated embedding as context. 
Layer widths and channel counts are those of the deployed configuration. 
The training of the flow involves learning the \emph{normalizing} transformation of the underlying posterior into the base density.}
  \label{fig:FlowArchitecture}
\end{figure*}
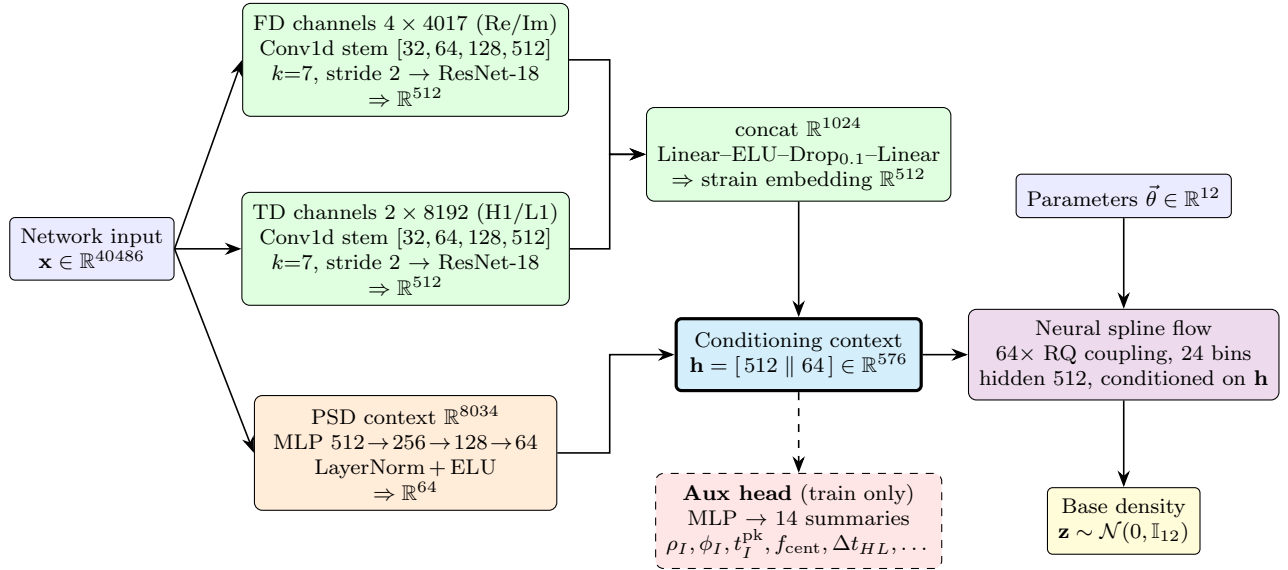

The normalizing flow uses rational quadratic spline transforms \cite{durkan2019neuralsplineflows}, with the spline parameters $\xi_i$ output by the per-layer conditioner networks described above. 
All the parameters have basic transforms to first regularize them into coordinates more suited for the flow (see, e.g., column~4 of Table~\ref{tab:FlowTrainingPriors}). 
Bounded parameters are further transformed via a sigmoid such that within the flow they are treated as unbounded parameters---a theoretical requirement for flows. 
Our specific flow architecture including the embedding, is detailed in Fig.~\ref{fig:FlowArchitecture}.
Other training aspects such as the hyperparameter of the neural networks involved and details of the priors are included in Appendix~\ref{sec:TrainingAppendix}. 

\section{Demonstration}
\label{sec:Results}
We now apply our conditional flow to LVK data to assess its performance.
We train on approximately $10^6$ waveforms, $10^4$ unique sky positions and $10^5$ unique noise realizations, taking 96 hours on an A100 NVIDIA GPU. 
We study the performance using both \emph{simulated} data with Gaussian noise and also simulated signals injected into ``off-source'' LIGO data where no known gravitational-wave signal is present.\footnote{The LIGO data for this study is taken from all available observing runs using the Gravitational-Wave Open Science Centre \cite[GWOSC;][]{Abbott2021GWOSC}.}
Real data is known to contain non-Gaussian artifacts not present in our training data \citep{Nuttall2018,Glanzer2024}.
By comparing these two datasets, we can assess how \texttt{BilbyFlow} responds to likelihood misspecification; for a broad discussion of misspecification in gravitational-wave astronomy, see Ref.~\cite{wmf}.

\begin{figure}[!ht]
     \centering
     \begin{subfigure}[b]{0.49\textwidth}
         \centering
         \includegraphics[width=\textwidth]{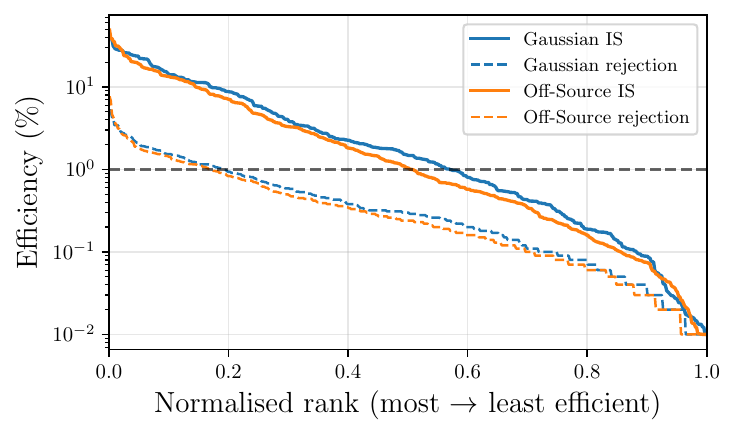}
     \end{subfigure}
    \caption{Comparison of importance-sampling efficiencies for simulated signals injected into Gaussian noise (blue) and injected into LIGO off-source data noise (orange). The equivalent efficiencies for rejection sampling are shown with the same colored dotted curves.
    We highlight the threshold for $1\%$ efficiency---which we regard as ``good''---with a dashed line.
    }
    \label{fig:SimulatedReweightingFig}
\end{figure}

Our results are shown in Fig.~\ref{fig:SimulatedReweightingFig}. 
In blue we show the efficiencies for {\emptyplaceholder{256}} events in simulated Gaussian noise.
In orange we show the efficiencies for the same number of events, injected into off-source LIGO segments.
For Gaussian noise, we meet our $\epsilon \geq 1\%$ goal for \paperchanges{\emptyplaceholder{57}}\% of the events.\footnote{
While we aim to improve \texttt{BilbyFlow} so that we can meet our $\epsilon>1\%$ target for 99\% of events, it is worth noting that, unlike traditional samplers, neural posterior estimators like \texttt{BilbyFlow} can be made embarrassingly parallel.
Thus, given sufficient computing power, they can still provide fast inference, even when the efficiency is very low.}
The median efficiency is {\emptyplaceholder{1.49}}\% and the worst efficiency is {\emptyplaceholder{0.01}}\%.
This shows that---for well behaved noise---\texttt{BilbyFlow} can perform inference about 10 times faster than traditional methods for most events.
For a small fraction of events, it performs comparatively poorly, but even then, it is no slower than traditional inference.
In orange we show the results for simulated signals in off-source LIGO data.
The efficiencies for pure Gaussian noise and off-source noise events are closely comparable.



\begin{figure*}[!ht]
    \centering
    \includegraphics[width=\linewidth]{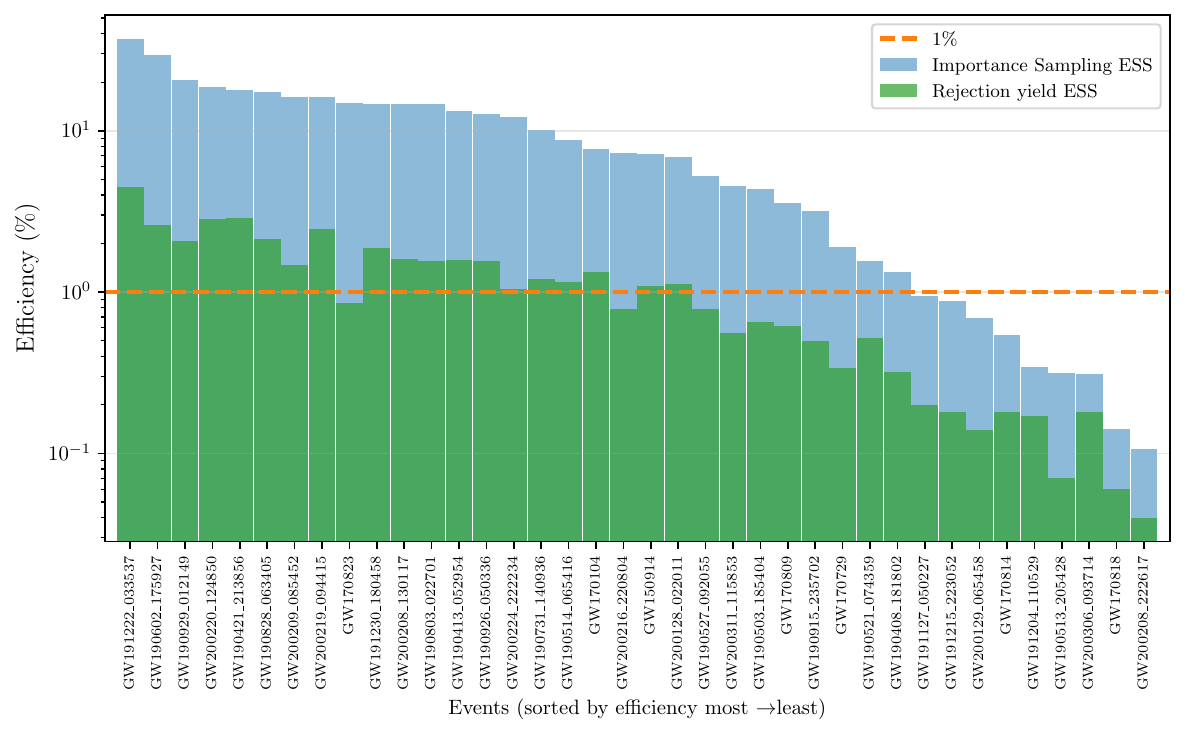}
    \caption{The ordered reweighting efficiencies for real event data segments taken from \texttt{GWOSC} \cite{Abbott2021GWOSC} for the GWTC-3 catalog. The orange dashed line represents the 1\% level reweighting efficiencies. All but \emptyplaceholder{nine} of the events \paperchanges{fall above} the 1\% level, corresponding to \paperchanges{23.7\%} of analyzed events.}
    \label{fig:RealDataEfficiencies}
\end{figure*}

We apply the flow to real event data taken from the GWTC-3 data release \cite{GWTC3_paper, GWTC_dataset}, and plot the efficiencies of the events in Fig.~\ref{fig:RealDataEfficiencies}. 
We meet our $\epsilon > 1\%$ goal for all but \emptyplaceholder{nine} of the 38 events.
The median efficiency is \emptyplaceholder{7}\%, achieving comparable results to \texttt{DINGO} \cite{Kofler_2026_DINGO_Transformers}.
The worst efficiency is \emptyplaceholder{0.11}\% (for \emptyplaceholder{GW200208\_222617}). This is notably higher than the synthetic data results, likely reflecting selection effects with detected events occupying regions where the flow performs well.
This may be due to the fact that in these regions signals are easier to observe, meaning the flow and embedding can more effectively find features in the data relating to the parameters of interest. In Appendix~\ref{sec:restricted_efficiencies} we perform some rough selection cuts to observe the efficiencies in these regions and observe an equalization of efficiencies with the on-source event data.
Using 16 cores, the computation times to get $10^4$ posterior samples for on-source events ranges from {\emptyplaceholder{$\unit[3]{mins}-\unit[35]{hr}$}} depending on the efficiency.

\begin{figure*}[!ht]
     \centering
     \begin{subfigure}[b]{0.42\linewidth}
         \centering
         \includegraphics[width=\textwidth, trim={0 0 3.5cm 0}, clip]{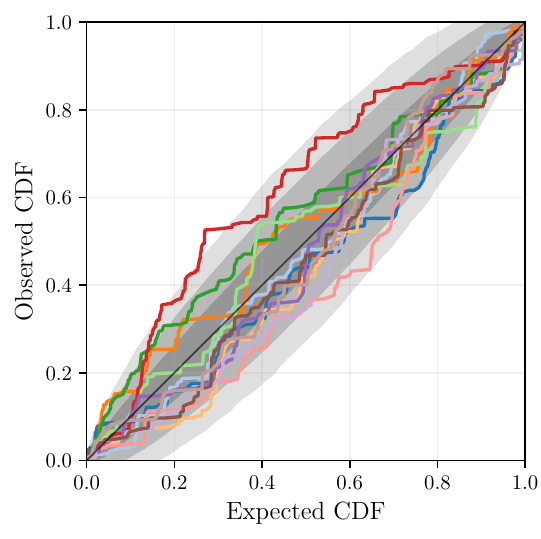}
     \end{subfigure}
     \begin{subfigure}[b]{0.55\linewidth}
         \centering
         \includegraphics[width=\textwidth, trim={0.65cm 0 0 0}, clip]{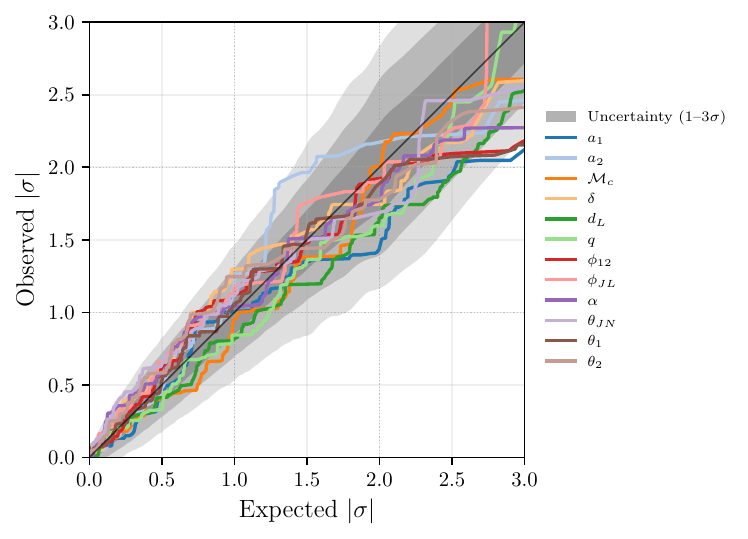}
     \end{subfigure}
     \begin{subfigure}[b]{0.42\linewidth}
         \centering
         \includegraphics[width=\textwidth, trim={0 0 3.5cm 0}, clip]{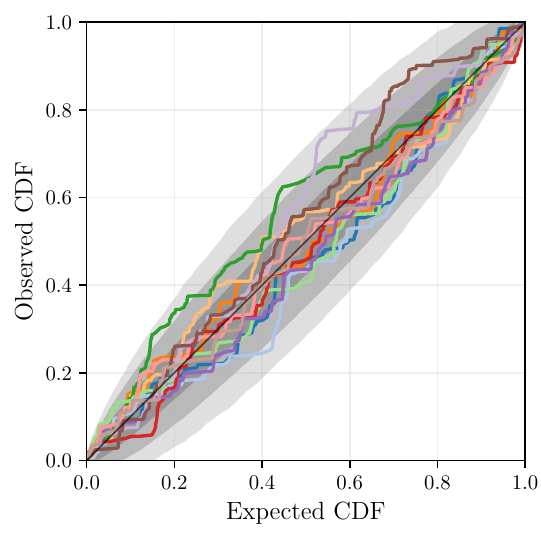}
     \end{subfigure}
     \begin{subfigure}[b]{0.55\linewidth}
         \centering
         \includegraphics[width=\textwidth, trim={0.65cm 0 0 0}, clip]{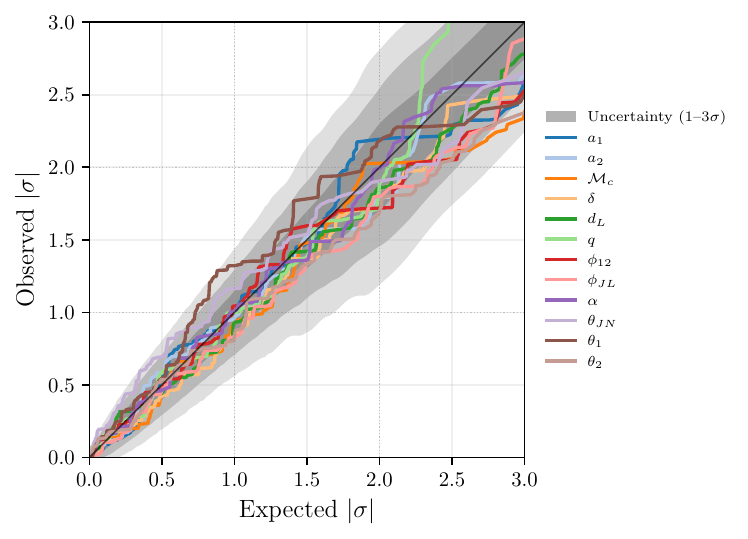}
     \end{subfigure}
    \caption{Calibration (pp) curves of the reweighted flow posterior in raw probability space (left column) and $|\sigma|$ space (right column) for simulated signals in Gaussian noise (top row, 512 injections) and in off-source LIGO data (bottom row, 512 injections).  The variance in these plots come from the finite number of simulations and true values used, and the finite number of reweighted posterior samples used to estimate the credibility contours, which is dominating here due to the mean ESS of 15 for these runs. A well-calibrated posterior follows the diagonal (dashed); curves below the diagonal indicate over-confident (too narrow) posteriors, while curves above indicate conservative (too broad) posteriors.  Shaded bands denote the 1, 2, and 3$\sigma$ expected scatter for the number of injections.
    The flow on both noise sources shows good calibration to $\sim\!3\sigma$, with deviations appearing in the tails indicating \paperchanges{over-confident} posterior width estimates in the tails of the distributions.
    } 
    \label{fig:Synthetic_Waveform_pp_plots}
\end{figure*}

We also generate pp-plots where we simulate a large number of events and note the smallest credibility contour in the subsequent posterior contains the true value that generated the data. In the simpler case of a standard one-dimensional normal distribution, this equates to recording the z-scores of the true values.\footnote{It is more standard to report the relevant probabilities instead of the z-scores, e.g., instead of $1\sigma$ one reports the fractional probability of $0.68$. But for evidence estimation we are particularly interested in the behavior in the tails of the distribution. Thus, we analyze the coverage of our flow approximations at $2\sigma$ and $3\sigma$ and beyond, which is difficult to see in a standard pp-plot.} 
The results of this for simulated data using Gaussian noise and off-source LIGO noise are shown in Fig.~\ref{fig:Synthetic_Waveform_pp_plots}.
The fact that the pp-plots are well-behaved for events in simulated Gaussian noise show that \texttt{BilbyFlow} produces the correct posterior distributions given our noise model.
The fact that the pp-plots show deviations from the desired behavior for events injected into off-source LIGO noise show that the widely-used Gaussian noise model is misspecified, and tends to produce \paperchanges{marginally over-confident} credible intervals in the tails of the distribution.

\begin{figure*}[!ht]
     \centering
     \begin{subfigure}[b]{0.49\linewidth}
         \centering
         \includegraphics[width=\textwidth]{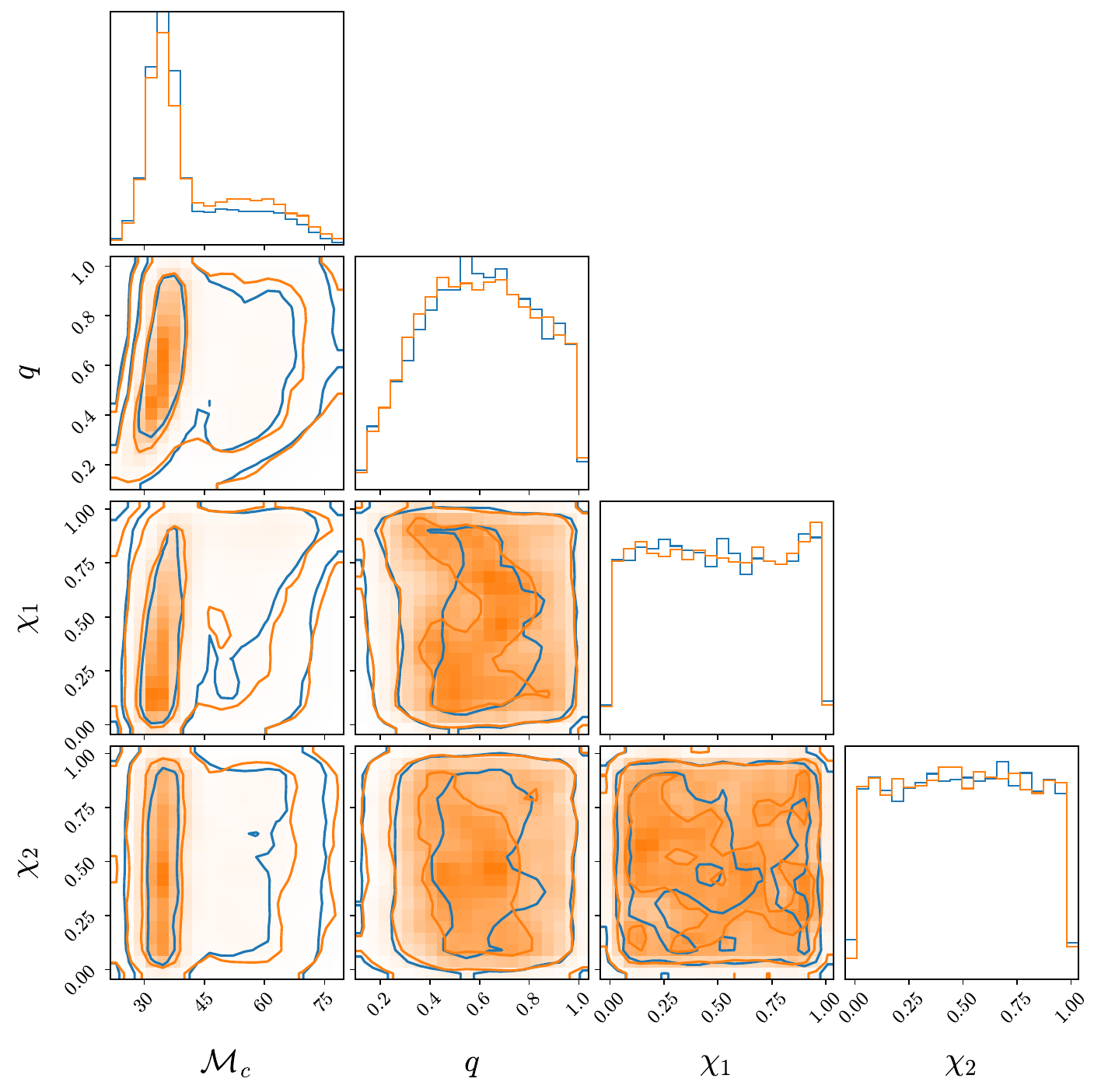}
         \caption{}
     \end{subfigure}
     \hfill
     \begin{subfigure}[b]{0.49\linewidth}
         \centering
         \includegraphics[width=\textwidth]{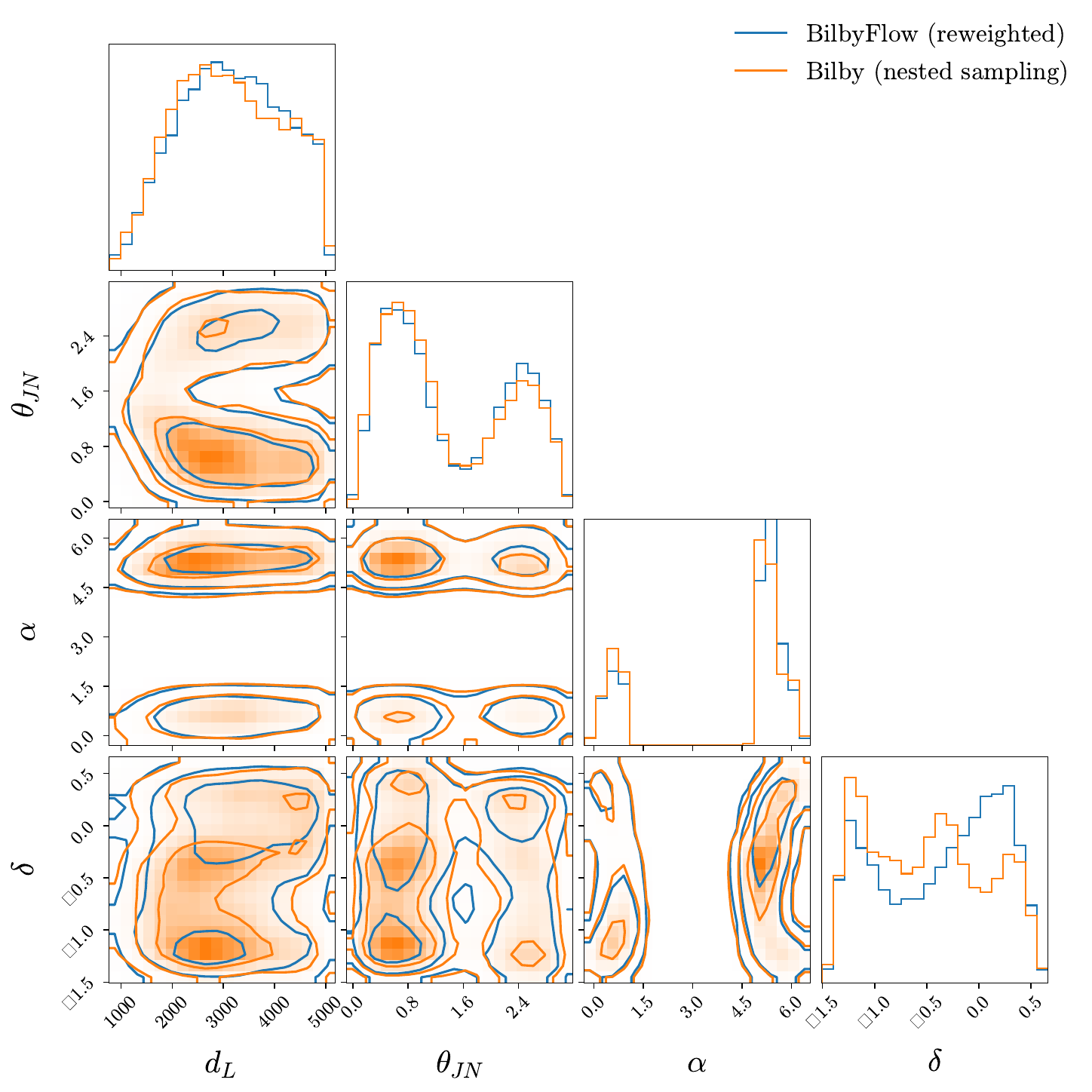}
         \caption{}
     \end{subfigure}
    \caption{
    A comparison of posterior credible intervals obtained with \texttt{BilbyFlow} (blue) to posterior credible intervals obtained with \texttt{Bilby} (orange) for the event GW200216\_220804. The left-hand plot (a) show intrinsic parameters: chirp mass $\mathcal{M}_c$, mass ratio $q$, and dimensionless spin $(\chi_1, \chi_2)$. The right hand plot (b) shows the extrinsic parameters: luminosity distance $d_L$, inclination angle $\theta_{JN}$, right ascension $\alpha$, and declination $\delta$.}
    
    \label{fig:BilbyFlow_vs_Bilby}
\end{figure*}

In Fig.~\ref{fig:BilbyFlow_vs_Bilby}, we compare posterior corner plots made with \texttt{BilbyFlow} and \texttt{Bilby} for the event GW200216\_220804. 
The close agreement shows that \texttt{BilbyFlow} is able to faithfully reproduce results from \texttt{Bilby}.
The \texttt{BilbyFlow} results are obtained in {\emptyplaceholder{30}} minutes on 16 cores. By comparison, a standard \texttt{Bilby} nested-sampling run for the same event and same number of cores takes approximately \emptyplaceholder{4} hours.

\section{Discussion and Conclusions}\label{sec:discussion}
We introduce \texttt{BilbyFlow}, a neural posterior estimation framework for the \texttt{Bilby} inference suite. 
\texttt{BilbyFlow} learns how to approximate the posterior distribution of compact binary parameters conditioned on gravitational-wave data.
Once trained, \texttt{BilbyFlow} can be used to generate $10^4$ posterior samples within {\emptyplaceholder{$\approx\unit[17]{min}$}} (median compute time with 16 cores).
This is a significant speed-up over traditional stochastic samplers, which take hours to days.
For \paperchanges{\emptyplaceholder{57\%}} of simulated events injected into Gaussian noise or off-segment noise, \texttt{BilbyFlow} meets our goal of an importance-sampling efficiency of $\gtrsim 1\%$.

As things stand, \texttt{BilbyFlow} is a useful tool for the majority of high-mass events (chirp mass $M_c\gtrsim10 M_\odot$) for which it was trained.
However, additional development is required for our longterm goal: to make \texttt{BilbyFlow} reliably fast for $\gtrsim 99\%$ of LVK events.
Here we discuss some of the improvements that are likely to be required to meet this new goal.

\textbf{Better flows.}
Improvements could include the use of flow matching \cite{flow_matching_guide_and_code} to allow for better utilization of known parameter symmetries and better scaling behaviors. This replaces the discrete coupling-layer transformation sequence with a learned ordinary differential equation, which scales better to high dimensions and larger dataset sizes.
Additionally, alternative training objectives (e.g., score matching or diffusion-based posteriors) could also provide better tail coverage where our efficiency losses concentrate.

\textbf{Better embedding architectures.}
The embedding network is the information bottleneck of the pipeline: the flow can only condition on what the embedding retains and structures well. 
The memory and time budget for the training is already nearing the limits of our currently available hardware.
As such, future works will need to utilize the data more effectively for example with JEPA-like (Joint Embedding Predictive Architecture) \cite{assran2023jepa} or transformer \cite{Kofler_2026_DINGO_Transformers} architectures, or expand training with distributed computing.
Because \texttt{BilbyFlow} is modular, either change is a drop-in replacement that does not impact the reweighting architecture.

\textbf{Per-event fine-tuning.}
The flow architecture provides a reasonable proposal for most events, and is immediately amortized to these events. 
However, we can treat the flow as an initialization for a variational approximation. 
Particularly for events with low reweighting efficiencies, we could run a short fine-tuning pass (as small as 10 steps) that updates the flow weights on the given set of event data.
The target is unchanged, so the reweighting guarantee is preserved, with fine-tuning only impacting proposal quality and increasing efficiency, not the correctness of the final samples.

\textbf{Training scale and diversity.}
The lack of a single dominant predictor of low efficiency (see Appendix~\ref{sec:efficiency_relationships}) suggests that the flow is uniformly under-trained rather than failing in a specific parameter regime. 
This is despite the flow showing early signs of \emph{over}-training when trained for longer on the current dataset.
This could indicate that the bottleneck is training data diversity rather than compute time.
Increasing the effective training diversity, via larger batch sizes or importance-weighted sampling that enrich underrepresented prior regions, would sharpen the proposal across the full parameter space and mitigate training issues involved with simply training for longer.
The prior-swap infrastructure\footnote{This is where raw flow proposal samples are initially reweighted to the physically-motivated prior.} already corrects any training-target prior mismatch introduced by such weighting. 
So these modifications would only impact the training loop without changing the reweighting pipeline.


\textbf{Low-mass events.}
The first version of \texttt{BilbyFlow} is focused on high-mass events with $M_c\gtrsim10 M_\odot$ because these events present a comparatively easier data-science problem than their lower-mass cousins.
Signals with lower masses contain more cycles in the LVK observing band.
As such, they produce more complicated structure in our data.
This complicated structure demands a more sophisticated neural net in order to build a model linking the data to the binary parameters.
In the future, these aspects can be included by first increasing the time window of the flow inputs and training on the wider mass range. 
If it is true that the data is inherently more complex and the NPE performs poorly due to this, one could increase the size of the embedding and flow.
Additionally, because \texttt{BilbyFlow} is modular, one could change either the embedding architecture (including transformers, e.g. \cite{Kofler_2026_DINGO_Transformers}) or flow (e.g. using flow matching).

\textbf{GPU acceleration.}
Where possible, \texttt{BilbyFlow} uses graphical processor units to speed up embarrassingly parallel calculations.
The current computational bottleneck is due to the fact that, during the reweighting step, we must evaluate the gravitational waveform on CPU.
However, it should be possible to ``\texttt{CUDA}-fy'' the gravitational waveform code so that the waveform can be evaluated on a GPU.
Once the waveforms have been ported over to \texttt{CUDA}, we expect that the reweighting compute time will be reduced by a factor of 10-100.
Since the overall run time is dominated by this step, we are hopeful that this could reduce inference times to $\unit[10]{sec}$ for the best events and to $\unit[2]{hours}$ for the worst.
Thus, we identify the creation of \texttt{CUDA} waveforms as a high-priority development project.

\textbf{Other applications.}
There may be other applications for \texttt{BilbyFlow} that make use of the ${\cal O}(\unit[1]{s})$ run time for initial sample generation.
For example, running \texttt{BilbyFlow} on data as it becomes available could potentially be used for low-latency detection and sky-map generation.
In this hypothetical pipeline, the more expensive reweighting step would only be triggered if a candidate signal was found in the initial flow samples.
Indeed, previous work has already demonstrated the possibility of doing rapid detection and sky-map estimation with dedicated neural nets; see, e.g., \citep{Chayan2019,Chayan2021,Chayan2023}.
Our point here is just that it might be possible to obtain similar predictions as a byproduct of a more general NPE pipeline.

We aspire for  \texttt{BilbyFlow} to become significantly more reliable, producing rapid results for close to 99\% of high-mass events.
The outlook is more uncertain for low-mass events as significant advances may be required in the deployment of our conditional normalizing flow.
\texttt{BilbyFlow} is open-source,\footnote{The code is available on GitHub at \href{https://github.com/LiamCPinchbeck/BilbyFlow}{LiamCPinchbeck/BilbyFlow}.} \texttt{pip}-installable, and inherits the modularity of the \texttt{Bilby} ecosystem: new waveform models, priors, and likelihoods become available to it without significant modification. With demonstrated reweighting efficiencies of $\gtrsim 1\%$ for the majority of GWTC-3 events and inference times of seconds to minutes, \texttt{BilbyFlow} provides a practical path towards the routine, rapid parameter estimation that current \emph{and} next-generation observatories will demand.

\subsection*{Acknowledgments}{We thank David Frazier for helpful discussions relating to robust SBI approaches. This work was performed on the OzSTAR national facility at Swinburne University of Technology. The OzSTAR program receives funding in part from the Astronomy National Collaborative Research Infrastructure Strategy (NCRIS) allocation provided by the Australian Government, and from the Victorian Higher Education State Investment Fund (VHESIF) provided by the Victorian Government. 
This material is based upon work supported by NSF's LIGO Laboratory which is a major facility fully funded by the National Science Foundation.
E.T. and P.D.L. are supported by ARC CE170100004, LE210100002, DP230103088, and CE230100016.
The research of C.B. is supported by ARC DP220100643, LE210100015 and LE250100010.}

\appendix

\section{Additional Training Information}
\label{sec:TrainingAppendix}

This appendix contains extra details \paperchanges{on} the training-side design choices that do not explicitly appear in the main body text but would be useful for reproducibility. 

\subsection{Bounded parameters and flow coordinates}

The normalising flow operates on standardised coordinates
$\hat\theta = (\theta - \mu_\theta)/\sigma_\theta$, where $\mu_\theta$ and
$\sigma_\theta$ are estimated from a representative training batch drawn
across the full prior range. We use the neural spline flow (NSF)
architecture~\citep{durkan2019neuralsplineflows} with rational-quadratic splines and linear
tails outside the finite spline domain $[-B,B]$, as implemented in the
\texttt{sbi} package\footnote{\paperchanges{The public package has migrated to the Zuko package for broader ease-of-use \cite{rozet2022zuko}.}} \citep{Tejero-Cantero2020}.

Parameters with compact support (e.g. spin magnitudes $\chi_i \in [0,0.99]$, mass ratio $q\in[0.125, 1]$) are handled by a sigmoid transform that maps $\mathbb{R}\to [-1, 1]$ in \paperchanges{the sampling direction; its inverse (a scaled logit) is applied in the normalizing direction.}
The Jacobian of each such transform (this and the following) is tracked explicitly and absorbed into the log-probability evaluation.

Two parameters use training proposals deliberately reshaped relative to the physical prior and corrected in the importance weights: the luminosity distance $d_L$ (log-uniform over $[100, 5000]$~Mpc, padded to $5200$~Mpc, to enrich coverage of loud nearby sources where efficiency is most sensitive) and optionally the chirp mass $\mathcal{M}_c$ (log-uniform, padded below the physical lower bound). In both cases the importance-sampling weights $w\propto\mathcal{L}(\theta|x)\pi_\text{phys}(\theta) / q_\text{train}(\theta|x;\varphi)$ correct the mismatch, and the Jacobians for the $\ln d_L$ coordinate changes are applied to the proposal density at evaluation time. Using `phys' to denote the physically motivated prior and $q_\text{train}(\theta|x;\varphi)$ to denote the density trained on the modified training proposals.

\subsection{Sky coordinate reparameterization}

The flow models the sky position in detector-frame coordinates $(\Delta t_\mathrm{HL},\, \phi_\mathrm{det})$ rather than the equatorial coordinates $(\alpha, \delta)$.
In this frame, $\Delta t_\mathrm{HL}$ is the differential arrival time between the LIGO Hanford and Livingston detectors and $\phi_\mathrm{det}$ is the azimuthal angle of the source in the detector plane. 
Both quantities are directly constrained by the data, whereas $(\alpha, \delta)$ couple to the data only through the time-dependent Earth-rotation matrix, creating a complicated and time-varying geometry that is harder for the flow to represent.

The conversion $(\Delta t_\mathrm{HL},\, \phi_\mathrm{det})\to(\alpha, \delta)$ is performed analytically at inference time given a reference GPS time $t_0$. 
Under the default target prior (`detector-uniform'), the flow's implicit prior in $(\Delta t_\mathrm{HL},\, \phi_\mathrm{det})$ matches the training distribution, and the constant Jacobian cancels in the importance weights. 
The isotropic sky prior is \paperchanges{handled via an explicit correction term} $\ln\pi(\alpha,\delta) - \ln|J|$, where $J =\partial(\Delta t_\mathrm{HL}, \phi_\mathrm{det}) / \partial(\alpha,\delta)$.

\subsection{Luminosity distance curriculum}

Training proceeds through a sequence of curriculum stages with increasing maximum luminosity distance $d_{L}^\text{max}$, as listed in Table.~\ref{tab:FlowTrainingPriors}. 
Each stage generates training data only from the sky bank entries satisfying $d_L\leq d_{L}^\text{max}$, so the effective SNR floor decreases progressively.
This curriculum both accelerates early learning on loud, information-rich signals that we have found to be difficult to constrain well, but also to create strain representations that do not expend capacity on noise. Before slowly increasing the level of noise to make the representations robust against different realizations and to larger noise levels which dominate the prior volume.

Each stage has its own learning rate and cosine-annealing schedule (see Table~\ref{tab:FlowTrainingPriors}). The standardization of $\hat\theta$ is computed once over the full prior range and held fixed across all stages; this avoids shocking the flow with abrupt rescaling at stage boundaries, at the cost of a mild scale mismatch during early stages where the effective $d_L$ range is narrower than the standardizer assumed.

\subsection{Auxiliary supervision}

An auxiliary regression head is attached to the shared embedding during training. This head is a three-layer MLP (hidden size 256) that predicts 14 noise-signal summary statistics from the embedding: per-detector matched-filter SNR, peak phase, peak time, and spectral centroid for each of the H1 and L1 detectors, plus the inter-detector time delay $\Delta t_\text{HL}$, SNR ratio between the detectors, and the phase difference.
The auxiliary loss $L_\text{aux}$ is annealed to zero over the course of each curriculum stage (reaching zero by a specific fraction $f_\text{anneal}$ of the stage).
The head is never used at inference time and does not affect the reweighting methods.

The purpose of the auxiliary head is to shape the embedding in early training by injecting a supervised signal that directly rewards physically informative feature directions. 
In the absence of such a signal, the only gradient shaping the embedding is the negative log-likelihood, which is initially weak (the flow has not yet learned to use the embedding) and risks the embedding collapsing onto a low-rank subspace. The extra loss term is included alongside the forward-KL loss from the flow, with $\lambda_\text{aux}$ on the order of 0.5 at the beginning of each curriculum stage,
\begin{align}
    \mathcal{L}_{\text{tot}} = L_\text{KL}(\varphi) + \lambda_\text{aux} L_\text{aux}.
\end{align}

\begin{table*}[!ht]
  \centering
  \caption{Priors for the parameters of the source model, the training proposal
    used to generate representative training datasets, the coordinate in which
    each parameter is represented inside the normalising flow, and the nature of
    that reparameterisation. The flow infers twelve parameters; the coalescence
    phase $\phi_c$, polarisation $\psi$, and coalescence time $t_c$ are not flow
    outputs but are semi-analytically marginalised$^{\S}$ in the importance-%
    reweighting target. Two parameters use proposals deliberately reshaped
    relative to the physical prior and corrected back by the importance weights
    $w \propto \mathcal{L}\,\pi_{\mathrm{phys}}/q$: the luminosity distance
    (log-uniform, padded to $5200$~Mpc, to enrich coverage of loud, nearby
    sources) and the chirp mass (log-uniform, padded below the physical lower
    bound). Flow coordinates are additionally standardised to zero mean and unit
    variance; bounded-support (sigmoid) transforms internal to the flow are not
    shown.}
    \vspace{0.5em}
    \label{tab:FlowTrainingPriors}
    \input{Tables_and_Diagrams/PriorTable}
    \vspace{0.6em}
\end{table*}

\section{Efficiency Relationships}
\label{sec:efficiency_relationships}

\begin{figure*}[!ht]
     \centering
     \begin{subfigure}[b]{0.98\linewidth}
         \centering
         \includegraphics[width=\linewidth, trim={0 0cm 0cm 0.cm}, clip]{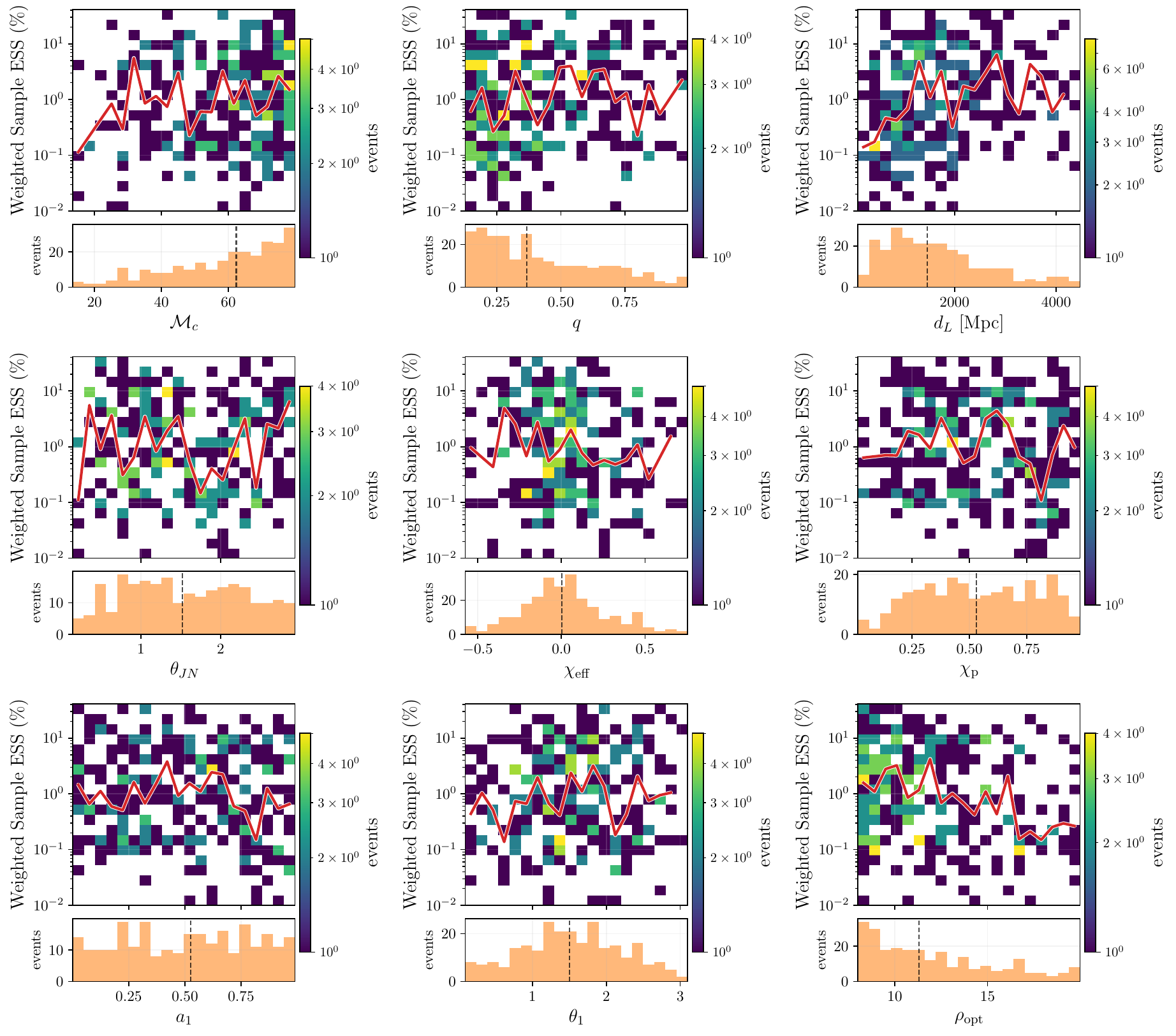}
     \end{subfigure}
    \caption{Figures showing histograms between the reweighting efficiencies of simulated injections in off-source LIGO noise segments with respect to some of the parameters of interest. This overlaid with a red curve which indicates the median efficiency for each parameter bin, and subsequently any broad relationships between the parameters and the reweighting efficiencies for the given events. The variance and overall behavior of the histograms and curves is biased by the priors used as part of the analysis, and hence they are shown in a 1D histogram below each main plot. The variable with the strongest relationship with the efficiencies is the chirp mass and optimal SNR, marginally indicating a decrease in efficiency as mass decreases and SNR increases. This relationships must be taken generously however, as the drop in efficiency is also in regions with fewer events and thus have a higher variance.}
    \label{fig:DataEfficiencyVsParams}
\end{figure*}

This appendix \paperchanges{examines} how source \paperchanges{parameters} correlate with reweighting efficiency. Fig.~\ref{fig:DataEfficiencyVsParams} shows per-event efficiency against several parameters of interest for synthetic injections in off-source LIGO noise. 
There is a mild relationship with the chirp mass and optimal SNR, showing that as the chirp mass decreases and SNR increases the efficiency decreases, but the efficiencies in the low mass and high SNR region have a large variance due to the sparsity of events there.
As such, there is no single parameter that has a \emph{strong} relationship with the efficiency: the flow is sub-optimally constraining uniformly across the parameter space. 
This indicates that the flow as a whole needs to relate the data to the parameters more strongly, or there is some high dimensional structure that cannot be seen on these effectively single dimensional plots.

\section{Restricted parameter space efficiencies}
\label{sec:restricted_efficiencies}

\begin{figure*}[!ht]
     \centering
     \begin{subfigure}[b]{0.45\linewidth}
         \centering
         \includegraphics[width=\linewidth, trim={0 0cm 0cm 0.cm}, clip]{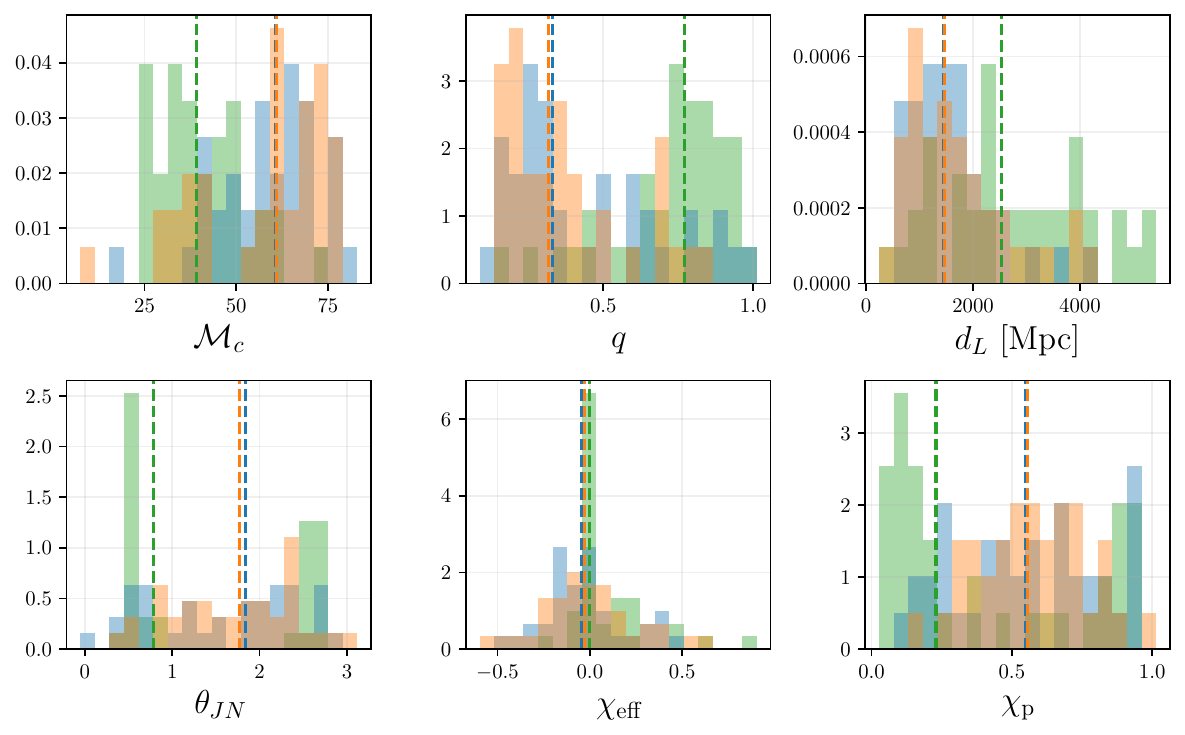}
         \caption{}
     \end{subfigure}
     \hfill
     \centering
     \begin{subfigure}[b]{0.51\linewidth}
         \centering
         \includegraphics[width=\linewidth, trim={0 0cm 0cm 0.cm}, clip]{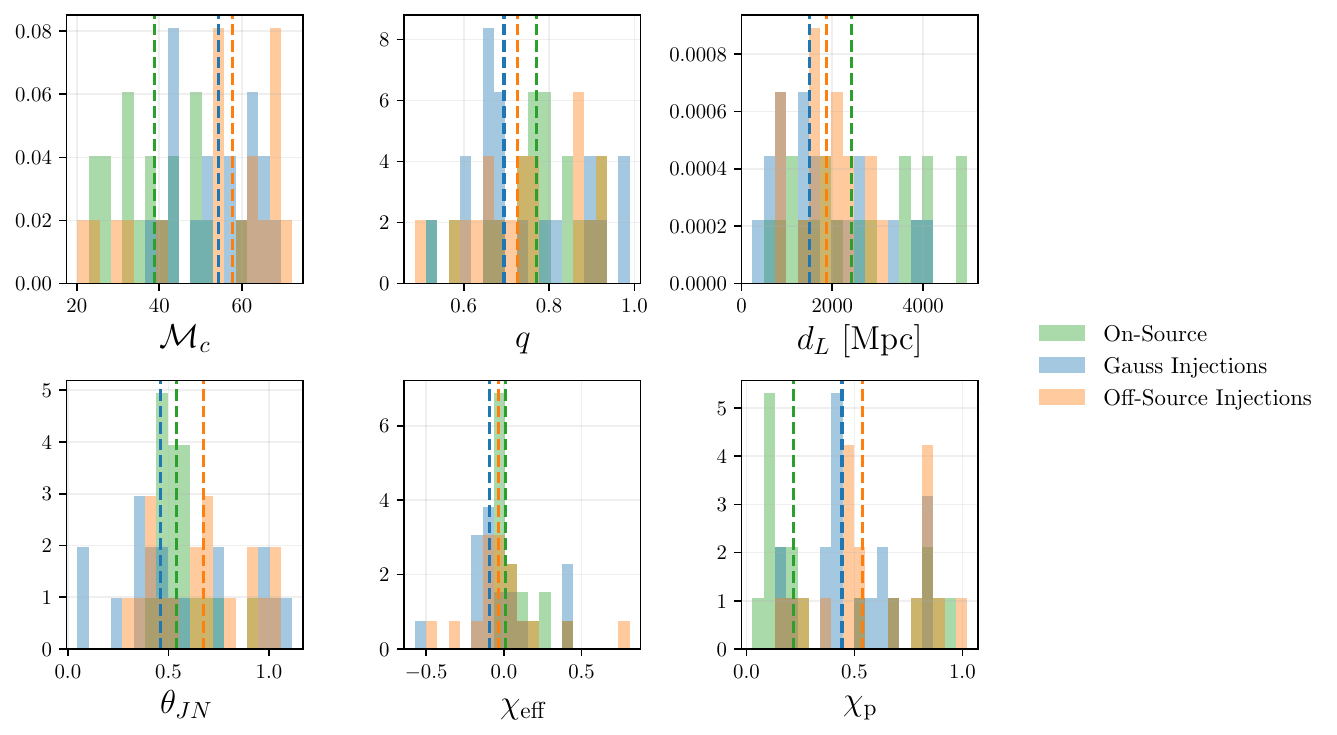}
         \caption{}
     \end{subfigure}
    \caption{\textbf{(a)} Histograms of parameters used as part of diagnostics with no selection cuts performed for on-source (green, using published posterior sample modes), waveform injections into Gaussian noise (blue), and waveform injections into off-segment noise (orange). The medians of each distribution are shown with dashed vertical lines. \textbf{(b)} Histograms of parameters used as part of diagnostics after selection cuts on the inclination angle ($0\rightarrow 1.1$) and mass ratio ($0.5\rightarrow 1$) have been performed.}
    \label{fig:parameter_selection_cuts}
\end{figure*}

As can be seen by comparing Fig.~\ref{fig:RealDataEfficiencies} and Fig.~\ref{fig:SimulatedReweightingFig}, the performance on synthetic data is notably worse than the on-source event data.
The most likely reason that this would occur is that the flow performs better in regions of parameter space that the real events inhabit.
We show a rudimentary selection cut which can be seen in Fig.~\ref{fig:parameter_selection_cuts}.
The primary selection cut was performed on the inclination angle, where the modes of which in published posterior samples exhibits a much stronger bimodality compared to that in the standard set of priors. 

\begin{figure*}[!ht]
     \centering
     \begin{subfigure}[b]{0.48\linewidth}
         \centering
         \includegraphics[width=\linewidth, trim={0 0cm 0cm 0.cm}, clip]{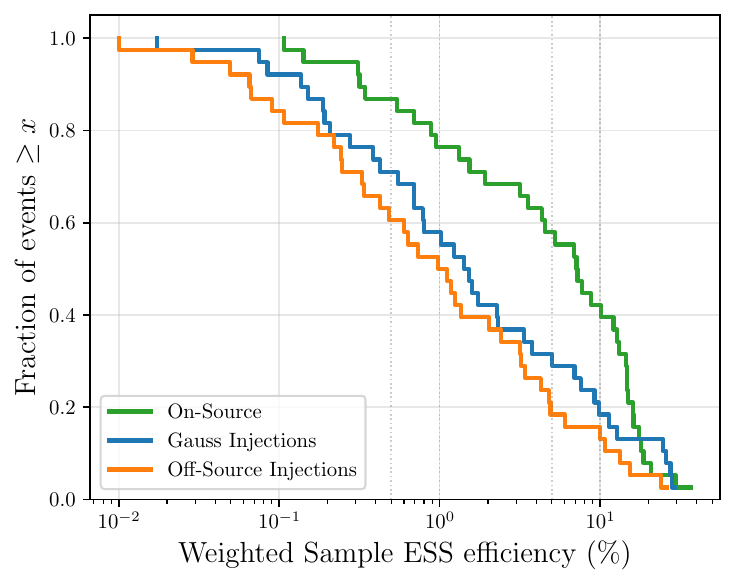}
         \caption{}
     \end{subfigure}
     \hfill
     \centering
     \begin{subfigure}[b]{0.48\linewidth}
         \centering
         \includegraphics[width=\linewidth, trim={0 0cm 0cm 0.cm}, clip]{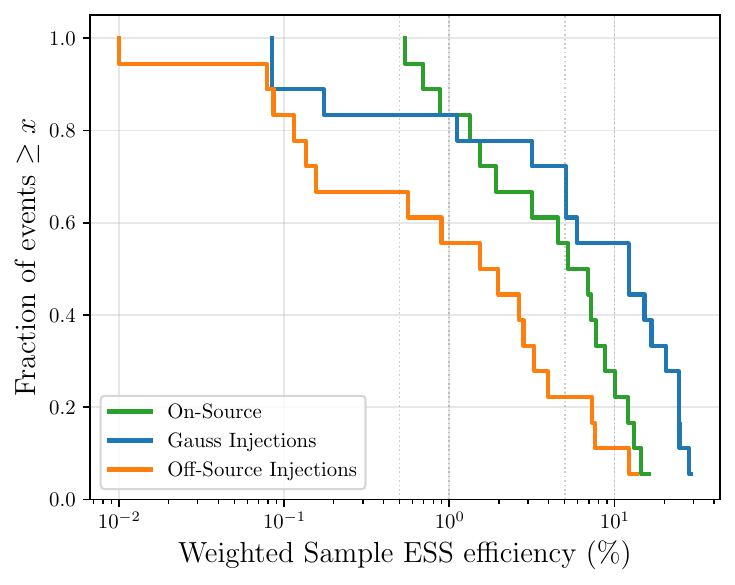}
         \caption{}
     \end{subfigure}
    \caption{\textbf{(a)} Survival curves showing the fraction of events that pass the weight efficiency threshold indicated on the horizontal axis for on-source (green), waveform injections into Gaussian noise (blue), and waveform injections into off-segment noise (orange). \textbf{(b)} Survival curves showing the fraction of events that pass the weight efficiency threshold indicated on the horizontal axis after selection cuts indicated in Fig.~\ref{fig:parameter_selection_cuts}. After the cuts the performance on synthetic waveform injections into pure Gaussian noise is comparable if not better than with on-source data. The performance on injection into off-segment noise only marginally changes indicating some secondary effect is reducing efficiencies for that data.}
    \label{fig:survival_selection_cuts}
\end{figure*}

In this restricted parameter space, we can more directly compare the performance of \texttt{BilbyFlow} between synthetic and on-source data. 
We show the effect of the selection cuts on efficiency survival curves in Fig.~\ref{fig:survival_selection_cuts}.
These curves show the fraction of events with efficiencies higher than those indicated on the horizontal axis.
After these selection cuts the performance of the framework between on-source data and data with synthetic injected waveforms injected into Gaussian noise are almost equivalent, showing roughly a factor of 4-10 improvement in the reweighting efficiencies on the synthetic data. 
This may be a selection bias where detected events are also those with clearer signals in the data, making feature extraction easier for the embedding and subsequently a better posterior representation from the flow.
Interestingly the performance on injections into off-segment data does not notably improve.
Although not confirmed, we suspect that this is due to another selection effect where the off-segment data used is more likely to have non-Gaussian noise leading to out-of-dataset errors in the framework.

\section{Further Information on the Reweighting Procedure}\label{sec:prior_swap}

The normalizing flow models the posterior over the 12 parameters detailed in the body. $\theta\in \mathbb{R}^{12}$. Three extrinsic parameters, $\lambda= (\phi_c, \psi, t_c)$, are not inferred by the flow; they are randomized during training so the flow learns the $\lambda$-marginal posterior $q(\theta|d)$ automatically. 
Consistency then requires that the importance-sampling target use the marginal likelihood (Eq.~\ref{eq:synth_marg}) rather than a point evaluation at fixed $\lambda$.
\begin{align}
    \mathcal{L}_{\textrm{marg}}(\theta) = \int \mathcal{L}(d\vert \theta, \lambda)\, \pi(\lambda)\, d\lambda\,,
    \label{eq:synth_marg}
\end{align}

The training proposal $\pi_\text{train}(\theta)$ (log-uniform in $d_L$) generally differs from the physical motivated target prior $\pi_\text{phys}(\theta)$ (e.g. $d_L^2$ power law). 
Before any likelihood evaluation, $N_\text{draw}$ flow samples are resampled via importance resampling with ratio $r(\theta) = \pi_\mathrm{phys}(\theta) / \pi_\mathrm{eff}(\theta)$. 

\begin{figure*}[!ht]
     \centering
     \begin{subfigure}[b]{0.49\linewidth}
         \centering
         \includegraphics[width=\textwidth]{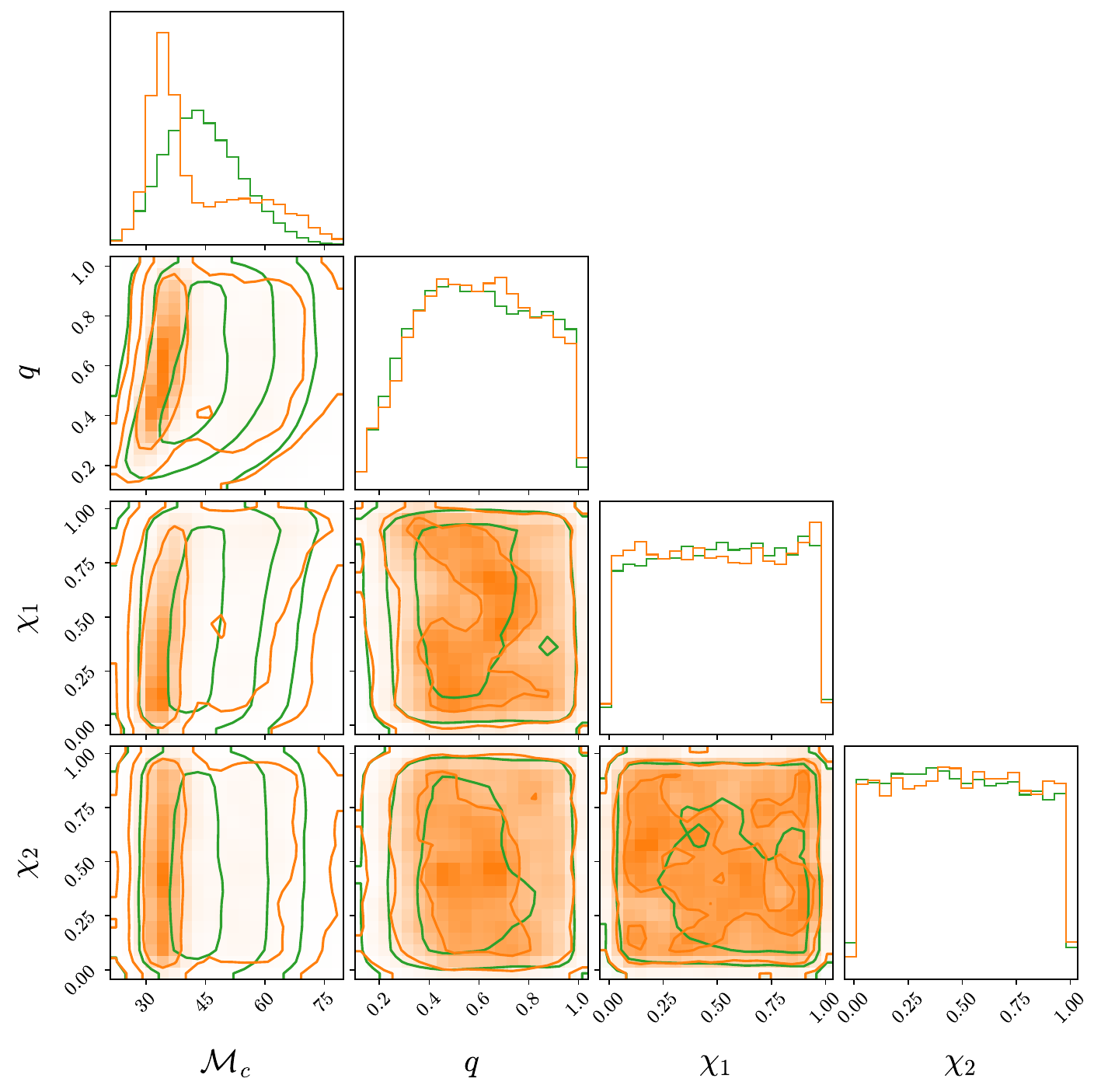}
         \caption{}
     \end{subfigure}
     \hfill
     \begin{subfigure}[b]{0.49\linewidth}
         \centering
         \includegraphics[width=\textwidth]{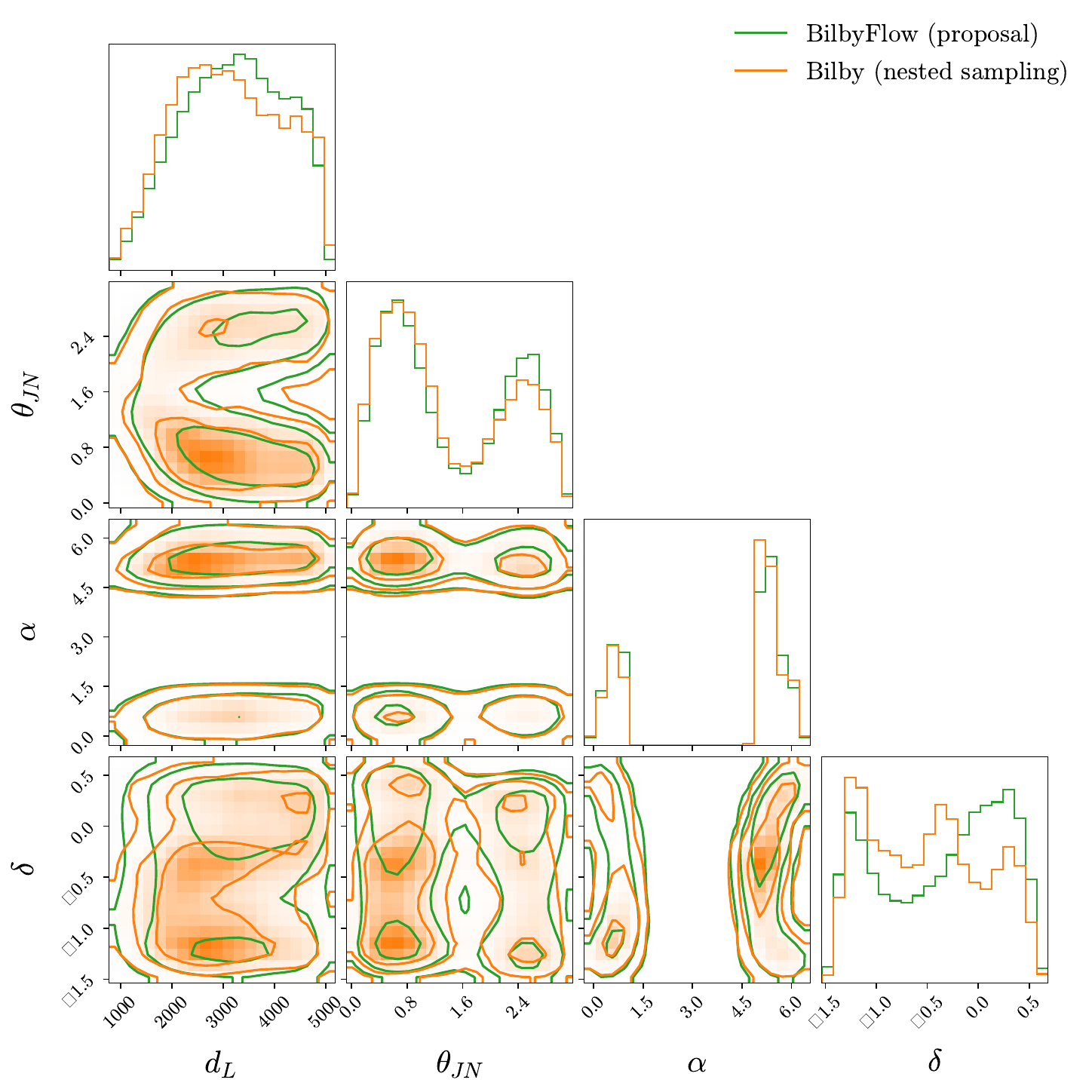}
         \caption{}
     \end{subfigure}
    \caption{A comparison of credible intervals obtained using \texttt{BilbyFlow}'s  posterior samples (blue) with ones obtained using proposal samples (orange) for the event GW200216\_220804. The left-hand plot (a) shows intrinsic parameters: chirp mass $\mathcal{M}_c$, mass ratio $q$, and dimensionless spins $(\chi_1, \chi_2)$. The right hand plot (b) shows extrinsic parameters: luminosity distance $d_L$,  inclination angle $\theta_{JN}$, right ascension $\alpha$, and declination $\delta$.
    }
    \label{fig:BilbyFlow_prior_vs_BilbyFlow_full}
\end{figure*}

\begin{figure*}[!ht]
     \centering
     \begin{subfigure}[b]{0.42\linewidth}
         \centering
         \includegraphics[width=\textwidth, trim={0 0 3.5cm 0}, clip]{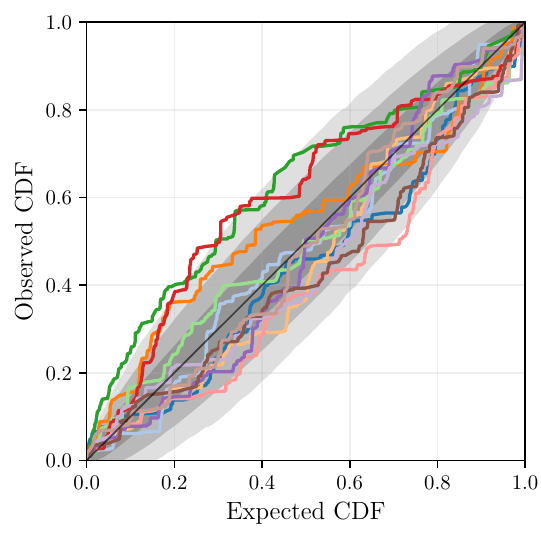}
     \end{subfigure}
     \begin{subfigure}[b]{0.55\linewidth}
         \centering
         \includegraphics[width=\textwidth, trim={0.65cm 0 0 0}, clip]{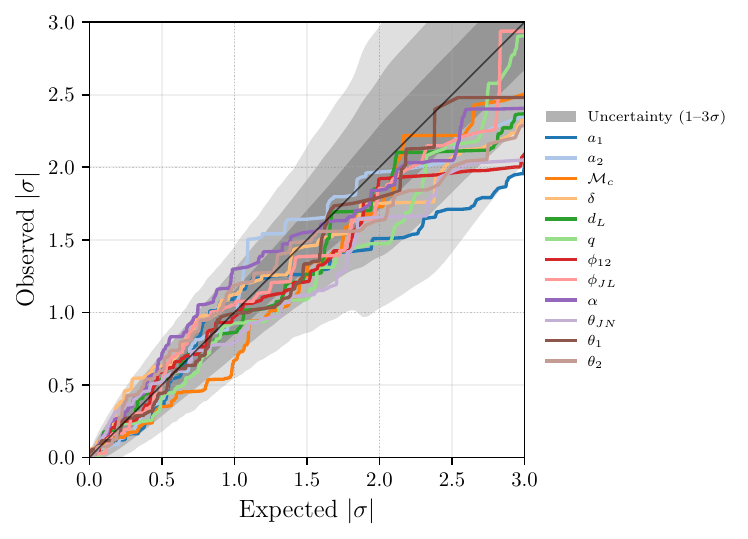}
     \end{subfigure}
     \begin{subfigure}[b]{0.42\linewidth}
         \centering
         \includegraphics[width=\textwidth, trim={0 0 3.5cm 0}, clip]{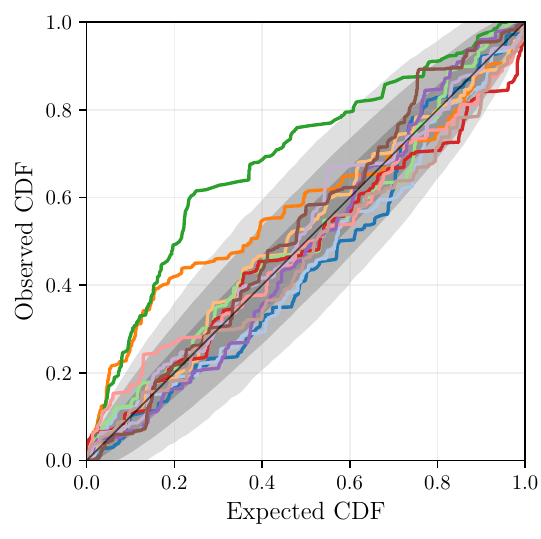}
     \end{subfigure}
     \begin{subfigure}[b]{0.55\linewidth}
         \centering
         \includegraphics[width=\textwidth, trim={0.65cm 0 0 0}, clip]{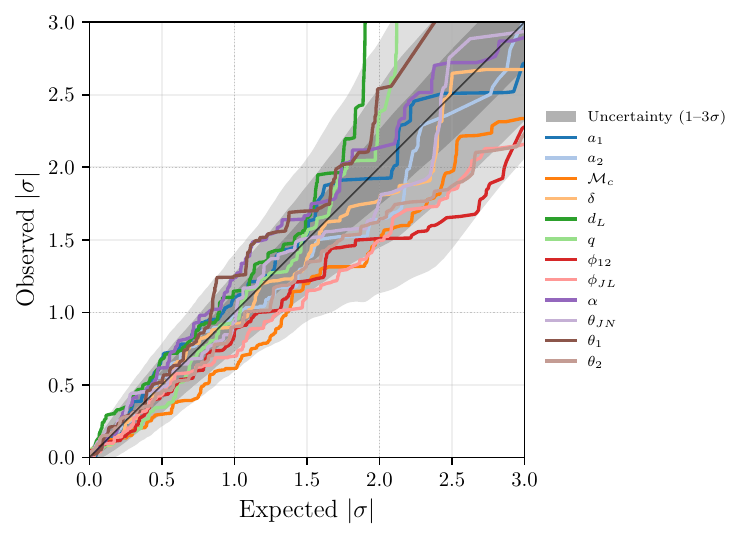}
     \end{subfigure}
    \caption{Calibration (pp) curves of the \texttt{BilbyFlow} posterior proposal samples in $|\sigma|$ space for simulated signals in Gaussian noise (left, 512 injections) and in off-source LIGO data (right, 512 injections).  The variance in these plots come from the finite number of simulations and true values used, and the finite number of posterior proposal samples used to estimate the credibility contours. A well-calibrated posterior follows the diagonal (dashed); curves below the diagonal indicate over-confident (too narrow) posteriors, while curves above indicate conservative (too broad) posteriors.  Shaded bands denote the 1, 2, and 3$\sigma$ expected scatter for the number of injections.
    The flow on both noise sources shows good calibration to $\sim\!2\sigma$, with deviations appearing in the tails indicating mild over-confidence in the extreme tails. Because the deviations are mild and confined beyond $\sim\!2\sigma$, the reweighted results remain reliable; nevertheless, over-confident proposal tails are the principal risk for importance-sampling bias, and motivate the training-diversity improvements discussed in the text.
    } 
    \label{fig:Synthetic_Waveform_raw_flow_pp_plots}
\end{figure*}

The resampled proposal density is $\tilde{q}(\theta) = q(\theta)\,r(\theta)/Z_r$, which replaces $q$ in weight evaluations. 
An oversampling factor of 10 ensures sufficient ESS (typically 30-80\%). 
The results of this stage of reweighting can be seen in Fig.~\ref{fig:BilbyFlow_prior_vs_BilbyFlow_full} compared against posterior draws from nested sampling using \texttt{dynesty}.
We can also compare the self-consistency of this stage, where we look at pp-plots relating to raw proposal samples from the flow, and whether they report consistent reported credibility levels. 
This is shown in Fig.~\ref{fig:Synthetic_Waveform_raw_flow_pp_plots}.
These plots indicate a marginal over-confidence in the tail regions, and under-confidence in the bulk probability mass region when applied to injections into off-segment noise.

The former indicates that the flow is having trouble capturing the tails of the distributions, possibly due to finite batch sizes and limited overall training bank examples.
The marginal under-confidence particularly in the luminosity distance implies that the flow or embedding is having trouble properly capturing details across the range of SNRs in the data.
In both cases the key direction of improvement is more diversified training as indicated by other tests above.

\paperchanges{The next reweighting stage then reweights to the XPHM \texttt{Bilby} likelihood, marginalizing over the nuisance parameters as in Eq.~\ref{eq:synth_marg} and Sec.~\ref{sec:subtleties}. The marginalization of the likelihood does not rely on the analytic identity that is exact only for (2,2)-dominated signals. The phase dependence of XPHM ($\ell \leq 4$) is multi-harmonic, $h(\varphi_c) = \Sigma_m C_m e^{im \phi_c}$ with $m = \{1,2,3,4\}$; we reconstruct the harmonic coefficients $C_m$ exactly via a DFT over $n = 5$ phase-sampled waveform evaluations (with an explicit anti-aliasing check) and marginalize the full multi-harmonic likelihood numerically over $(\phi_c, \psi, t_c)$ grids with local refinement, following the synthetic-phase approach of \cite{Dax_2023}. The remaining discretization is in the $\psi/t_c$ grids only, is mode-agnostic, and is validated end-to-end by the PP tests. }

\bibliography{refs}


\end{document}

%% file: Tables_and_Diagrams/general_conditional_flow_architecture.tex

\begin{tikzpicture}[
  font=\footnotesize,
  nfblk/.style ={draw,rounded corners=4pt,minimum width=10mm,minimum height=8mm,
                 fill=cyan!12,thick,inner sep=2pt},
  nfgrk/.style ={draw,rounded corners=4pt,minimum width=10mm,minimum height=8mm,
                 fill=green!8,thick,inner sep=2pt},
  nfblkf/.style={draw,rounded corners=4pt,minimum width=10mm,minimum height=8mm,
                 fill=orange!28,thick,inner sep=2pt},
  nfembed/.style={draw,rounded corners=5pt,fill=black!80,text=white,align=center,
                 inner sep=5pt,minimum height=8mm},
  nfnn/.style  ={draw,rounded corners=4pt,fill=blue!12,align=center,inner sep=3pt},
  nfnng/.style ={draw,dashed,rounded corners=4pt,fill=blue!5,align=center,
                 inner sep=3pt,text=black!55},
  nfxi/.style  ={draw,rounded corners=4pt,fill=violet!16,align=center,inner sep=2pt,
                 minimum width=8mm},
  nfftr/.style ={draw,rounded corners=3pt,fill=gray!15,align=center,inner sep=2pt,
                 minimum size=7mm},
  nfxdat/.style={draw,rounded corners=4pt,fill=red!16,align=center,inner sep=8pt},
  nfch/.style  ={-{Stealth[length=2.6mm]},line width=1.5pt,gray!60},
  nffl/.style  ={-{Stealth[length=2mm]},semithick},
  nfflg/.style ={-{Stealth[length=1.8mm]},semithick,dashed,teal!55!black},
  nfbus/.style ={line width=1.4pt,teal!65!black},
  nfdens/.style={draw,dashed,rounded corners=3pt,minimum size=23mm,inner sep=0pt},
]
\node[nfxdat]  (x)   at (1.7,5.2) {Data ${d}$};
\node[nfembed] (enc) at (5.5,5.2) {Embedding network};
\draw[nffl] (x) -- (enc);

\draw[nfbus] (enc.south) -- (5.5,4.1);
\draw[nfbus] (1.3,4.1) -- (10.1,4.1);
\node[teal!65!black] at (1.85,4.38) {context $S$};

\node[nfgrk]  (z0)  at (0,0)     {$z_0$};
\node[nfftr]  (f1)  at (1.3,0)   {$f_1$};
\node[nfblk]  (z1)  at (2.6,0)   {$z_1$};
\node         (cd)  at (3.7,0)   {$\cdots$};
\node[nfblk]  (zim) at (4.8,0)   {$z_{i-1}$};
\node[nfftr]  (fi)  at (6.1,0)   {$f_i$};
\node[nfblk]  (zi)  at (7.4,0)   {$z_i$};
\node         (cd2) at (8.5,0)   {$\cdots$};
\node[nfftr]  (fK)  at (10.1,0)  {$f_K$};
\node[nfblkf] (zK)  at (11.4,0)  {$z_K$};
\node[right=1mm of zK] {$\approx\theta$};

\foreach \a/\b in {z0/f1,f1/z1,z1/cd,cd/zim,zim/fi,fi/zi,zi/cd2,cd2/fK,fK/zK}
  \draw[nfch] (\a) -- (\b);

\node[nfnn] (nn1) at (1.3,2.7)   {$\mathrm{NN}_1$\\(conditioner)};
\node[nfxi] (x1)  at (1.3,1.35)  {$\xi_1$};
\node[nfnn] (nni) at (6.1,2.7)   {$\mathrm{NN}_i$\\(conditioner)};
\node[nfxi] (xii) at (6.1,1.35)  {$\xi_i$};
\node[nfnn] (nnK) at (10.1,2.7)  {$\mathrm{NN}_K$\\(conditioner)};
\node[nfxi] (xiK) at (10.1,1.35) {$\xi_K$};
\node[nfnng] (nng1) at (3.4,2.7) {$\vdots$};
\node[nfnng] (nng2) at (8.0,2.7) {$\vdots$};

\draw[nffl,teal!65!black] (1.3,4.1)  -- (nn1.north);
\draw[nffl,teal!65!black] (6.1,4.1)  -- (nni.north);
\draw[nffl,teal!65!black] (10.1,4.1) -- (nnK.north);
\draw[nfflg] (3.4,4.1) -- (nng1.north);
\draw[nfflg] (8.0,4.1) -- (nng2.north);

\foreach \n/\xv/\fv in {nn1/x1/f1,nni/xii/fi,nnK/xiK/fK}{
  \draw[nffl] (\n) -- (\xv);
  \draw[nffl] (\xv) -- (\fv);
}
\draw[nffl,gray!70] (z0.north)  to[out=90,in=-90] (nn1.south west);
\draw[nffl,gray!70] (z1.north)  to[out=90,in=-90] (nng1.south west);
\draw[nffl,gray!70] (zim.north) to[out=90,in=-90] (nni.south west);
\draw[nffl,gray!70] (zi.north)  to[out=90,in=-90] (nng2.south west);
\draw[nffl,gray!70] (cd2.north) to[out=90,in=-90] (nnK.south west);

\node[gray!95,font=\scriptsize,below=0.5mm of fi] {$z_i=f_i(z_{i-1};\xi_i(S))$};

\node[nfdens] (D0) at (0,-2.2)    {};
\node[nfdens] (Di) at (7.4,-2.2)  {};
\node[nfdens] (DK) at (11.4,-2.2) {};

\foreach \cx in {0,7.4,11.4}{
  \draw[-{Stealth[length=1.2mm]}] (\cx-1.05,-2.7)  -- (\cx+1.05,-2.7);
  \draw[-{Stealth[length=1.2mm]}] (\cx-0.88,-2.78) -- (\cx-0.88,-1.25);
}

\begin{scope}[shift={(0,-2.7)}]
  \draw[purple,thick,smooth,samples=60,variable=\x,domain=-0.95:0.95]
    plot ({\x},{1.25*exp(-4*\x*\x)});
\end{scope}
\begin{scope}[shift={(7.4,-2.7)}]
  \draw[purple,thick,smooth,samples=80,variable=\x,domain=-0.95:0.95]
    plot ({\x},{1.00*exp(-8*(\x-0.35)*(\x-0.35))
               +1.10*exp(-7*(\x+0.30)*(\x+0.30))});
\end{scope}
\begin{scope}[shift={(11.4,-2.7)}]
  \draw[purple,thick,smooth,samples=100,variable=\x,domain=-0.95:0.95]
    plot ({\x},{0.75*exp(-12*(\x-0.50)*(\x-0.50))
               +1.15*exp(-14*(\x-0.06)*(\x-0.06))
               +0.65*exp(-12*(\x+0.45)*(\x+0.45))});
\end{scope}

\node[below=0.5mm of D0,font=\small] {$z_0\sim p_0(z_0)$};
\node[below=0.5mm of Di,font=\small] {$p_i(z_i\mid{d})$};
\node[below=0.5mm of DK,font=\small] {$p(\theta\mid{d})$};
\end{tikzpicture}

%% file: Tables_and_Diagrams/datagen_diagram_plain_english.tex

\begin{tikzpicture}[
  font=\scriptsize,
  W/.style     ={text width=70mm},
  theta/.style ={draw=blue!70,fill=blue!16,rounded corners=2pt,align=center,
                 inner sep=3pt,minimum height=6mm,text width=70mm},
  bank/.style  ={draw=blue!55,fill=blue!10,rounded corners=2pt,align=center,
                 inner sep=3pt,minimum height=6mm,text width=20mm},
  proj/.style  ={draw=teal!60,fill=teal!10,rounded corners=2pt,align=center,
                 inner sep=3pt,minimum height=6mm,W},
  noise/.style ={draw=orange!70,fill=orange!15,rounded corners=2pt,align=center,
                 inner sep=3pt,minimum height=6mm,W},
  chan/.style  ={draw=violet!55,fill=violet!10,rounded corners=2pt,align=center,
                 inner sep=3pt,minimum height=6mm,W},
  ctxs/.style  ={draw=magenta!50,fill=magenta!9,rounded corners=2pt,align=center,
                 inner sep=3pt,minimum height=6mm,W},
  outp/.style  ={draw=black!75,line width=0.7pt,fill=yellow!30,rounded corners=2pt,
                 align=center,inner sep=3pt,minimum height=6mm,W},
  aux/.style   ={draw=green!45!black,fill=green!12,rounded corners=2pt,align=center,
                 inner sep=3pt,minimum height=6mm,W,dashed},
  ar/.style    ={-{Stealth[length=2mm]},draw=black!60,semithick},
  arA/.style   ={-{Stealth[length=2mm]},draw=green!40!black,semithick,dashed},
  lbl/.style   ={font=\tiny,text=black!55,align=center},
]

\node[theta] (th) at (0,0)
  {Draw $\vec\theta$: intrinsic, sky, distance\;
   $\rightarrow\;\ln\mathcal{M},\,\ln d_L,\,\Delta t_{HL},\,\phi_{\mathrm{det}},\dots$};

\node[bank] (wf)  at (-24.5mm,-1.15) {Waveform bank\\{\tiny$h_+,h_\times$}};
\node[bank] (sky) at (0,-1.15)       {Sky bank\\{\tiny$F_{+,\times},\Delta t_d$}};
\node[bank] (psd) at (24.5mm,-1.15)  {PSD bank\\{\tiny$S_d(f)$, off-source}};

\node[proj]  (proj)  at (0,-2.35) {Project on H1/L1, scale to $d_L$, window,
                                   whiten by $b_d=\sqrt{4\Delta f S_d}$};
\node[noise] (noise) at (0,-3.3) {Add coloured Gaussian noise
                                   $w\sim\mathcal{CN}(0,1)$ from $S_d(f)$};
\node[chan]  (ch)    at (0,-4.2) {Channels: $\mathrm{Re}\,x_d,\ \mathrm{Im}\,x_d$,
                                   whitened TD};
\node[outp]  (x)     at (0,-5.10) {Training pair $(\mathbf{x},\vec\theta)$};
\node[aux]   (aux)   at (0,-6.05) {Auxiliary targets (training only):
                                   $\ln\rho_d,\,t^{\mathrm{peak}}_d,\,\ln f^{\mathrm{cent}}_d,\,
                                   \ln(\rho_{\mathrm H}/\rho_{\mathrm L}),\,\Delta\phi_{\mathrm{HL}}$};

\draw[ar] (th.south) -- ++(0,-2mm) -| (wf.north);
\draw[ar] (th.south) -- ++(0,-2mm) -| (sky.north);
\draw[ar] (wf.south)  -- ++(0,-3mm) -| (proj.north);
\draw[ar] (sky.south) -- ++(0,-3mm) -- (proj.north);
\draw[ar] (psd.south) -- ++(0,-3mm) -| (proj.north);
\draw[ar] (psd.east)  -- ++(4mm,0) |- (noise.east);
\draw[ar] (proj)  -- (noise);
\draw[ar] (noise) -- (ch);
\draw[ar] (ch)    -- (x);
\draw[arA] (proj.west) -- ++(-4mm,0) |- (aux.west);
\end{tikzpicture}

%% file: Tables_and_Diagrams/flow_architecture.tex

\begin{tikzpicture}[
  font=\footnotesize,
  fio/.style ={draw,rounded corners=2pt,fill=blue!8,align=center,inner sep=4pt},
  fenc/.style={draw,rounded corners=3pt,fill=green!12,align=center,inner sep=4pt,
               minimum width=40mm},
  fmlp/.style={draw,rounded corners=3pt,fill=orange!15,align=center,inner sep=4pt,
               minimum width=40mm},
  fctx/.style={draw,very thick,rounded corners=2pt,fill=cyan!15,align=center,inner sep=5pt},
  fflw/.style={draw,rounded corners=3pt,fill=violet!14,align=center,inner sep=4pt},
  fbas/.style={draw,rounded corners=2pt,fill=yellow!18,align=center,inner sep=4pt},
  faux/.style={draw,dashed,rounded corners=3pt,fill=red!10,align=center,inner sep=4pt},
  fa/.style  ={-{Stealth[length=2.2mm]},semithick},
  fad/.style ={-{Stealth[length=2.2mm]},semithick,dashed},
]

\node[fio] (x) at (1.15,0) {Network input\\$\mathbf{x}\in\mathbb{R}^{40486}$};

\node[fenc] (fd)  at (5.3, 2.5)  {FD channels $4\times4017$ (Re/Im)\\
                                  Conv1d stem $[32,64,128,512]$\\
                                  $k{=}7$, stride 2 $\to$ ResNet-18\\
                                  $\Rightarrow\mathbb{R}^{512}$};
\node[fenc] (td)  at (5.3, 0)    {TD channels $2\times8192$ (H1/L1)\\
                                  Conv1d stem $[32,64,128,512]$\\
                                  $k{=}7$, stride 2 $\to$ ResNet-18\\
                                  $\Rightarrow\mathbb{R}^{512}$};
\node[fmlp] (psd) at (5.3,-2.7)  {PSD context $\mathbb{R}^{8034}$\\
                                  MLP $512\!\to\!256\!\to\!128\!\to\!64$\\
                                  LayerNorm\,+\,ELU\\
                                  $\Rightarrow\mathbb{R}^{64}$};

\node[fenc] (head) at (10.5, 1.25) {concat $\mathbb{R}^{1024}$\\
                                   Linear--ELU--Drop$_{0.1}$--Linear\\
                                   $\Rightarrow$ strain embedding $\mathbb{R}^{512}$};
\node[fctx] (h)    at (10.5,-1.4)  {Conditioning context\\
                                   $\mathbf{h}=[\,512\;\|\;64\,]\in\mathbb{R}^{576}$};
\node[faux] (aux)  at (10.5,-3.6)  {\textbf{Aux head} (train only)\\
                                   MLP $\to$ 14 summaries\\
                                   $\rho_I,\phi_I,t^{\mathrm{pk}}_I,f_{\mathrm{cent}},\Delta t_{HL},\dots$};

\node[fio]  (theta) at (14.8, 0.7)  {Parameters $\vec\theta\in\mathbb{R}^{12}$};
\node[fflw] (flow)  at (14.8,-1.4)  {Neural spline flow\\
                                     $64\times$ RQ coupling, 24 bins\\
                                     hidden 512, conditioned on $\mathbf{h}$};
\node[fbas] (z)     at (14.8,-3.6)  {Base density\\$\mathbf{z}\sim\mathcal{N}(0,\mathbb{I}_{12})$};

\draw[fa] (x.east) -- (fd.west);
\draw[fa] (x.east) -- (td.west);
\draw[fa] (x.east) -- (psd.west);

\draw[fa] (fd.east)  -- (8., 2.5)  |- (head.west);
\draw[fa] (td.east)  -- (8., 0)    |- (head.west);
\draw[fa] (psd.east) -- (8., -2.7) |- (h.west);

\draw[fa]  (head.south) -- (h.north);
\draw[fad] (h.south)    -- (aux.north);
\draw[fa]  (h.east)     -- (flow.west);
\draw[fa]  (theta.south) -- (flow.north);
\draw[fa]  (flow.south)  -- (z.north);
\end{tikzpicture}

%% file: Tables_and_Diagrams/PriorTable.tex
  \makebox[\textwidth][c]{%
  \begin{tabular}{llll}
    \toprule
    Parameter                              & Training proposal                       & Flow coordinate          & Reparameterization \\
    \midrule
    Chirp mass $\mathcal{M}$ ($M_\odot$)   & Uniform, $[9.5,\,80]$   & $\mathcal{M}$        & --- \\
    Mass ratio $q$                         & Uniform in $m_{1,2}$, $[0.125,\,1]$     & $q$                      & --- \\
    Luminosity distance $d_L$ (Mpc)        & Log-uniform, $[100,\,5000]$ & $\ln d_L$                & log distance \\
    Inclination $\theta_{JN}$              & Sine, $[0,\,\pi]$                       & $\theta_{JN}$            & --- \\
    Right ascension $\alpha$               & Uniform, $[0,\,2\pi]$                   & $\Delta t_{HL}$ & H1--L1 time delay \\
    Declination $\delta$                   & Cosine, $[-\pi/2,\,\pi/2]$              & $\phi_{\mathrm{det}}$ & detector azimuth \\
    Spin magnitude $a_1$                   & Uniform, $[0,\,0.99]$                   & $a_1$                    & --- \\
    Spin magnitude $a_2$                   & Uniform, $[0,\,0.99]$                   & $a_2$                    & --- \\
    Spin tilt $\theta_1$                   & Sine, $[0,\,\pi]$                       & $\theta_1$               & --- \\
    Spin tilt $\theta_2$                   & Sine, $[0,\,\pi]$                       & $\theta_2$               & --- \\
    Spin azimuth $\phi_{12}$               & Uniform, $[0,\,2\pi]$                   & $\phi_{12}$              & --- \\
    Spin azimuth $\phi_{JL}$               & Uniform, $[0,\,2\pi]$                   & $\phi_{JL}$              & --- \\
    Coalescence time $t_c$ (s)             & Uniform, $[-0.11,\,0.11]$               & ---             &  \\
    Polarisation $\psi$                    & Uniform, $[0,\,\pi]$                    & ---             &  \\
    Phase $\phi_c$                         & Uniform, $[0,\,2\pi]$                   & ---             &  \\
    \bottomrule
  \end{tabular}}
